\documentclass[10pt,final,doublecolumn]{IEEEtran}

\usepackage{graphicx}
\usepackage{epsfig,algorithmicx,algpseudocode}
\usepackage{amsmath}
\usepackage{amssymb}
\usepackage{subfigure}
\usepackage{wrapfig}
\usepackage{times,color}
\usepackage{cite,footnote,xspace,syntonly,algorithm,bm}
\input{mysymbol.sty}
\newcommand{\caliHm}{{\cal H}_{\cal M} }

\newcommand{\caliHn}{{\cal H}_{\cal N} }
\newcommand{\caliHp}{{\cal H}_{\cal P} }
\newcommand{\caliF}{{\cal F}}
\newcommand{\caliM}{{\cal M}}

\newcommand{\caliN}{{\cal N}}
\newcommand{\caliP}{{\cal P}}

\newcommand{\sumr}{\sum_{r=1}^{R}}
\newcommand{\sumn}{\sum_{n=1}^{N}}
\newcommand{\summ}{\sum_{m=1}^{M}}
\newcommand{\sump}{\sum_{p=1}^{P}}

\newcommand{\E}{{\mathbb{E}}}

\newcommand{\trace}{{\textrm{Tr}}}

\newcommand{\krprod}{\odot}

\newcommand{\boX}{{\mathbf{\bar X}}}
\newcommand{\boQ}{{\mathbf{\bar Q}}}

\newcommand{\boB}{{\mathbf{\bar B}}}
\newcommand{\boC}{{\mathbf{\bar C}}}
\newcommand{\bQ}{{\bf Q}}

\newcommand{\bhatz}{{\mathbf{\hat Z}}}
\newcommand{\bhatZ}{\underline{\mathbf{\hat{Z}}}}

\newcommand{\bZ}{{\bf Z}}

\newcommand{\bB}{{\bf B}}
\newcommand{\bI}{{\bf I}}
\newcommand{\bR}{{\bf R}}

\newcommand{\bX}{{\bf X}}
\newcommand{\bD}{{\bf D}}
\newcommand{\Diag}{{\textrm{diag}}}
\newcommand{\bK}{{\bf K}}
\newcommand{\bA}{{\bf A}}
\newcommand{\bC}{{\bf C}}
\newcommand{\bk}{{\bm k}}

\newcommand{\bDelta}{{\bm \Delta}}

\newcommand{\bbept}{{\mathbf{e}^T_p}}

\newcommand{\XX}{\underline{\mathbf{ X}}}

\newcommand{\DD}{\underline{\bm{ \Delta}}}
\newcommand{\ZZ}{\underline{\mathbf{ Z}}}
\newcommand{\OO}{\underline{\mathbf{ O}}}
\newcommand{\hadprod}{{\circledast}}

\newcommand{\mR}{{\mathbb {R}}}

\long\def\symbolfootnote[#1]#2{\begingroup
\def\thefootnote{\fnsymbol{footnote}}
\footnote[#1]{#2}\endgroup} \psfull

\begin{document}
%--------------------------------------------The First Page---------------------------------------------------------------
\title{\huge Rank regularization and Bayesian  inference\\ for  tensor completion and extrapolation$^\dag$}

\author{{\it Juan Andr\'es Bazerque, Gonzalo~Mateos, and Georgios~B.~Giannakis~(contact author)$^\ast$}}

\markboth{IEEE TRANSACTIONS ON SIGNAL PROCESSING (SUBMITTED)}
\maketitle \maketitle \symbolfootnote[0]{$\dag$ This work was supported by
MURI (AFOSR FA9550-10-1-0567) grant. Part of the paper appeared in the
{\it Proc. of IEEE Workshop on Statistical Signal Processing}, Ann Arbor, USA, August 5-8, 2012.}
\symbolfootnote[0]{$\ast$ The authors are with the Dept. of
ECE and the Digital Technology Center, University of Minnesota,
200 Union Street SE, Minneapolis, MN 55455. Tel/fax:
(612)626-7781/625-2002; Emails:
\texttt{\{bazer002,mate0058,georgios\}@umn.edu}}

\markboth{IEEE TRANSACTIONS ON SIGNAL PROCESSING (SUBMITTED)}
\maketitle \maketitle

\thispagestyle{empty}\addtocounter{page}{-1}
\begin{abstract}
A novel regularizer of  the PARAFAC decomposition factors
capturing the tensor's rank is proposed in this paper, as the key enabler for
completion of three-way data arrays with missing entries. Set in a Bayesian framework,
the tensor completion method incorporates prior information to enhance its
smoothing and prediction capabilities. This probabilistic approach can
naturally accommodate general models for the data distribution, lending itself to various fitting criteria
that yield optimum estimates in the maximum-a-posteriori sense.
In particular, two algorithms are devised for Gaussian- and Poisson-distributed data, that minimize the rank-regularized
least-squares error and Kullback-Leibler divergence, respectively. The proposed technique is able to
recover the ``ground-truth'' tensor rank when tested on synthetic data, and to
complete  brain  imaging and yeast gene expression datasets with $50\%$ and $15 \%$ of
missing entries respectively, resulting in  recovery errors at $-10$dB and $-15$dB.
\end{abstract}

\vspace*{-5pt}
%\newpage
% for peerreview papers, inserts a page break and creates the second title.
% Will be ignored for other modes.
%\IEEEpeerreviewmaketitle
%\newpage
%-----------------------------------The Main Body-------------------------------------------------------------

\section{Introduction}
Imputation of missing data is a basic task arising in various Big Data
applications as diverse  as medical imaging \cite{GRY11}, bioinformatics \cite{ABB00}, as well
as social and computer networking \cite{kolda_poisson,MMG12}. The key idea
rendering recovery feasible is  the ``regularity'' present among missing and available
data. Low rank is an attribute capturing this regularity, and can be readily exploited
when data are organized in a matrix.
A natural approach to  \textit{low-rank matrix completion} problem
is  minimizing the rank of a target matrix, subject to a constraint
on the error in fitting the observed entries \cite{CP10}. Since rank minimization
is generally  NP-hard \cite{VB96},  the nuclear norm has been advocated recently as a convex
surrogate to the rank~\cite{F02}. Beyond  tractability, nuclear-norm
minimization enjoys good performance both in theory as well as in practice~\cite{CP10}.

The goal of this paper is imputation of missing entries of tensors (also known as multi-way
arrays), which are high-order generalizations of matrices frequently
encountered in chemometrics, medical imaging, and
networking~\cite{kolda_tutorial,czpa2009book}. Leveraging the low-rank structure
for tensor completion is challenging, since even computing the
tensor rank is NP-hard \cite{rankNPhard}. Defining a nuclear norm surrogate is not
obvious either, since singular values as defined by the Tucker
decomposition are not generally related with the rank. Traditional
approaches to finding low-dimensional representations of tensors
include unfolding the multi-way data  and applying matrix
factorizations such as the singular-value decomposition (SVD) \cite{CS09,THK11,ABB00} or, employing the parallel factor
(PARAFAC) decomposition~\cite{tBS02,jk77laa}. In the context
of tensor completion, an approach falling under the first category
can be found in~\cite{GRY11}, while imputation using  PARAFAC  was
dealt with in~\cite{kolda_completion}.

The imputation approach presented in this paper builds on a novel regularizer
accounting for the tensor rank, that relies on redefining the matrix nuclear norm
in terms of its low-rank factors. The contribution is two-fold. First, it is established that the low-rank
inducing property of the regularizer carries over to tensors by promoting sparsity
in the  factors of  the tensor's PARAFAC decomposition. In passing, this analysis
allows for drawing a neat connection with the atomic-norm in~\cite{CRPW12}.
The second contribution is the incorporation of prior information, with a Bayesian approach
that endows  tensor completion  with extra smoothing and prediction
capabilities.  A parallel analysis in the context of reproducing kernel Hilbert spaces
(RKHS) further explains these acquired capabilities, provides an alternative means of
obtaining the prior information, and establishes a useful connection with collaborative
filtering approaches \cite{ABEV09} when reduced to the matrix case.

While least-squares (LS) is typically utilized as the fitting criterion for matrix and
tensor completion, implicitly assuming Gaussian data, the adopted probabilistic
framework  supports the incorporation of   alternative data models.  Targeting count
processes  available in the form of network traffic data, genome sequencing, and social
media interactions, which are  modeled  as  Poisson distributed, the maximum a
posteriori (MAP) estimator is expressed in terms of the Kullback-Leibler (K-L) divergence~\cite{kolda_poisson}.
%
%Two algorithms are developed, namely the LRTI algorithm regularized LS imputation, and the LRPTI for  regularized KL imputation. Based on the a block successive upperbound minimization approach, they offer guaranteed convergence to a stationary point. Numerical test with both synthetic and real data corroborate the capability of these two algorithms to approximate the missing entries of a tensor by regularizing their ranks.

The remainder of the paper is organized as follows.  Section \ref{sec:preliminaries}
offers the necessary background on nuclear-norm regularization for matrices,  the PARAFAC
decomposition, and the  definition of tensor rank. Section \ref{sec:rank_tensors} presents the
tensor completion problem, establishing the low-rank  inducing property of the proposed regularization.
Prior information is incorporated in  Section \ref{sec:Bayesian}, with  Bayesian and RKHS formulations of
the tensor imputation method, leading to the low-rank tensor-imputation (LRTI) algorithm.
Section \ref{sec:poisson} develops the  method  for Poisson tensor data, and redesigns the algorithm
to minimize the rank-regularized K-L divergence. Finally, Section \ref{sec:numerical_tests} presents numerical
tests carried out on synthetic and real data, including expression levels in yeast, and brain magnetic
resonance images (MRI). Conclusions are drawn in Section VII, while most technical details are deferred to the Appendix.

The notation adopted throughout  includes bold lowercase and capital letters for vectors $\mathbf a$
and matrices $\mathbf A$, respectively, with superscript $T$ denoting transposition. Tensors are underlined
as e.g., $\XX$, and their slices carry a subscript as in $\mathbf X_p$; see also Fig. \ref{fig:regalo}. Both the matrix and tensor Frobenius
norms are represented by $\|\cdot\|_F$. Symbols $\otimes$,  $\krprod,$ $\hadprod$,  and $\circ,$
denote the  Kroneker, Kathri-Rao, Hadamard (entry-wise), and outer product, respectively.

\section{Preliminaries}\label{sec:preliminaries}

\subsection{Nuclear-norm minimization for matrix completion}
\label{ssec:nuclear}
Low-rank approximation is a popular method for
estimating missing values of a matrix $\bZ\in\mathbb{R}^{N\times M}$, which capitalizes
on ``regularities'' across the data \cite{F02}.
For the imputation to be feasible, a binding assumption
that relates the available entries with the missing ones is required.
An alternative is to postulate that  $\mathbf Z$
has low rank $R\ll\min(N,M)$. The problem of finding
matrix $\bhatz$ with rank not exceeding $R$, which approximates $\mathbf Z$
in the given entries specified by a binary
matrix $\bDelta\in\{0,1\}^{N\times M}$, can be formulated as
\begin{equation}
\bhatz=\arg \min_{\mathbf X}
\|{\bf(Z-X)}\hadprod {\bDelta}\|_F^2\;\;\; {\rm s.\ to\ \; rank }
(\mathbf X) \leq {R} \;.
\label{lowrank}
\end{equation}
The low-rank property of matrix $\mathbf{X}$ implies that the vector
$\mathbf s(\mathbf{X})$ of its singular values is sparse. Hence, the rank
constraint is equivalent to $\|\mathbf s({\bf X})\|_0\leq R,$
where the $\ell_0$-(pseudo)norm $\|\cdot\|_0$ equals the number of nonzero
entries of its vector argument.

%\footnote{The term ``zero-norm'' is
%widely used, even if it is not strictly a norm since it does not
%satisfy linearity.}
Aiming at a convex relaxation of the NP-hard
problem \eqref{lowrank}, one can leverage recent advances in compressive
sampling \cite{F02} and surrogate the $\ell_0$-norm
with the $\ell_1$-norm, which here equals the nuclear
norm of $\bf X$ defined as $\|{\bf X}\|_*:=\|{\bf s}({\bf X})\|_1$.
With this relaxation, the Lagrangian counterpart of \eqref{lowrank}
is
\begin{equation}
\bhatz=\arg \min_{\mathbf X}\frac{1}{2} \|{\bf
(Z-X)}\hadprod \bDelta\|_F^2+\mu\|\bf X\|_* \label{nuclearnorm}
\end{equation}
where $\mu\geq 0$ is  a rank-controlling parameter.
%Several  iterative
%algorithms have been proposed to solve \eqref{nuclearnorm}, and are
%effective in tackling low- to medium-size matrix completion
%problems; see e.g.,~\cite{?}. However, most
%algorithms require computation of singular values per iteration and
%become prohibitively expensive when dealing with high-dimensional
%data. To solve \eqref{nuclearnorm} efficiently, consider the following
Problem \eqref{nuclearnorm} can be further transformed by considering the following characterization of the nuclear norm~\cite{SRJ05}
\begin{equation}
\|\mathbf X\|_*=\min_{\{\mathbf{B},\mathbf{C}\}}{\frac{1}{2} (\|\mathbf B\|_F^2+\|\mathbf
C\|_F^2)}\quad
\textrm{s. to }~\mathbf{X}=\mathbf{B}\mathbf{C}^T. %,\: \mathbf C\in \mathbb R^{N\times
%r},\ \mathbf B\in \mathbb R^{N_g\times r} .
\label{nuclear_norm_decomposition}
\end{equation}
For an arbitrary matrix $\bX$ with SVD $\mathbf{X}=\mathbf{U}\bm{\Sigma}\mathbf{V}^T$, the minimum in
\eqref{nuclear_norm_decomposition} is attained for $\mathbf{B}=\mathbf{U}\bm{\Sigma}^{1/2}$
and $\mathbf{C}=\mathbf{V}\bm{\Sigma}^{1/2}$. The optimization in
\eqref{nuclear_norm_decomposition} is over all possible bilinear factorizations
of $\bX$, so that the number of columns of $\mathbf{B}$ and
$\mathbf{C}$ is also a variable.
Building on \eqref{nuclear_norm_decomposition}, one can
arrive at the following equivalent reformulation
of \eqref{nuclearnorm}~\cite{MMG12}
\begin{align}
\nonumber \bhatz'=&\arg \min_{\{\mathbf{X},\mathbf{B},\mathbf{C}\}}
\frac{1}{2}\|{\bf(Z-X)}\hadprod \bDelta\|_F^2+\frac{\mu}{2}(\|\mathbf
B\|_F^2+\|\mathbf C\|_F^2) \\&\quad{\rm s. \: to\;}\mathbf{X}=\mathbf B\mathbf{C}^T. \label{nuclear_norm_lr}
\end{align}
The equivalence implies that by finding the global minimum of
\eqref{nuclear_norm_lr},  one
can recover the optimal solution of \eqref{nuclearnorm}.
However, since \eqref{nuclear_norm_lr} is
\textit{nonconvex}, it may have multiple stationary points. Interestingly,
the next result provides conditions for these stationary points to be
globally optimal (parts a) and  b) are proved in the Appendix, while the proof for c) can be found in \cite{MMG12}.)

\begin{proposition}\label{prop:prop_matrix}
Problems \eqref{nuclearnorm} and \eqref{nuclear_norm_lr} are equivalent, in the sense that:
\begin{itemize}
\item[a)] global minima coincide: $\bhatz=\bhatz'$;
\item[b)] all local minima of \eqref{nuclear_norm_lr} are globally optimal; and,
\item[c)] stationary points $\bX$ of \eqref{nuclear_norm_lr}  satisfying $\|(\bX-\bZ)\hadprod \bDelta\|_2\leq \mu$ are globally optimal.
\end{itemize}
\end{proposition}

This result plays a critical role in this paper, as the Frobenius-norm regularization for controlling the rank in \eqref{nuclear_norm_lr},
will be useful to obtain its tensor counterparts in Section \ref{sec:rank_tensors}.
%This form of the low-rank estimator will be considered henceforth.
%
\subsection{PARAFAC decomposition}
The PARAFAC decomposition of a tensor
$\XX \in \mR^{M\times N\times P}$ is at the heart of the
proposed imputation method, since it offers a means to define its
 rank~\cite{tBS02,jk77laa}. Given $R \in \mathbb N$,
consider matrices $\bA\in \mR^{N\times R}$, $\bB\in \mR^{M\times R}$,
and $\bC\in \mR^{P\times R}$, such that
\begin{equation}
\XX(m,n,p)=\sum_{r=1}^R\bA(m,r)\bB(n,r)\bC(p,r).\label{threewayproduct}
\end{equation}
The rank of $\XX$ is  the minimum
value of $R$ for which this decomposition is possible. For $R^*:=\textrm{rank}(\XX)$,
the PARAFAC decomposition is given by the corresponding factor matrices
$\{\bA,\bB,\bC\}$ (all with $R^*$ columns), so that \eqref{threewayproduct} holds with $R=R^*$.
\begin{figure}[t]
  \centerline{\epsfig{figure=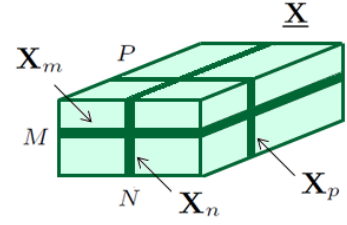,width= 0.5\linewidth}}
  \vspace{-.3cm}
\caption{Tensor slices along the row, column, and tube dimensions.}\vspace{-.3cm}
\label{fig:regalo}
\end{figure}

To appreciate why the aforementioned rank definition is natural, rewrite \eqref{threewayproduct}
as $\XX=\sum_{r=1}^R \mathbf a_{r}\circ \mathbf b_{r} \circ \mathbf c_{r}$,
where $\mathbf a_{r}$,  $\mathbf b_{r}$, and $\mathbf c_{r}$
represent the $r$-th columns of $\bA$, $\bB$, and $\bC$,
respectively; and the outer products $\OO_r:=\mathbf
a_{r}\circ\mathbf b_{r}\circ\mathbf c_{r}\in\mR^{M\times N\times P}$ have entries $\OO_r(m,n,p):=\bA(m,r)\bB(n,r)\bC(p,r).$
The rank of a tensor is thus the minimum number of outer products
(rank one factors) required to represent the tensor. It is not uncommon
to adopt an equivalent normalized representation
\begin{equation}
\XX=\sum_{r=1}^R \mathbf a_{r}\circ \mathbf b_{r} \circ \mathbf c_{r}=\sum_{r=1}^R \gamma_r (\mathbf u_{r}\circ \mathbf v_{r} \circ \mathbf w_{r})\label{normalized_outerproduct}
\end{equation}
by defining unit-norm vectors $\mathbf u_r:=\mathbf a_r/\|\mathbf a_r\|$,
$\mathbf v_r:=\mathbf b_r/\|\mathbf b_r\|$, $\mathbf w_r:= \mathbf c_r/\| \mathbf c_r\|$,
and weights  $\gamma_r:=\|\mathbf a_r\|\| \mathbf b_r\|\| \mathbf c_r\|$, $r=1,\ldots, R$.

Let  $\bX_p,\ p=1,\ldots,P$ denote the $p$-th slice of $\XX$ along its third (tube) dimension, such that
$\bX_p(m,n):=\XX(m,n,p)$; see Fig. \ref{fig:regalo}. The following compact form of the PARAFAC
decomposition in terms of  slice factorizations will be used in the sequel
\begin{equation}
\bX_p=\bA \Diag\left[\bbept \bC\right] \bB,\ ~~ p=1,\ldots,P\label{parafac}
\end{equation}
where the diagonal matrix $\Diag\left[\mathbf u\right]$ has the vector
$\mathbf u$ on its diagonal, and $\bbept$ is the $p$-th row of the $P\times P$ identity matrix.
The PARAFAC decomposition is symmetric [cf. \eqref{threewayproduct}],
and one can also write $\bX_m=\bB \Diag\left[{\bf e}_m^T \bA\right] \bC$,
or, $\bX_n=\bC \Diag\left[{\bf e}_n^T \bB\right] \bA$ in terms of slices
along the first (row), or, second (column) dimensions.

\section{Rank regularization for tensors}
\label{sec:rank_tensors}
Generalizing the nuclear-norm regularization technique  \eqref{nuclearnorm}
from low-rank matrix to tensor completion is not
straightforward, since  singular values of a tensor (given
by the Tucker decomposition) are not related
to the  rank~\cite{kolda_tutorial}. Fortunately, the Frobenius-norm regularization outlined in Section
\ref{ssec:nuclear}  offers a viable option for low-rank tensor completion
under the PARAFAC model, by solving
\begin{align}\label{tensor_approximation}
\bhatZ\hspace{-.05cm}:=\hspace{-.1cm}\arg \hspace{-.8cm}\min_{\{\XX,\bA,\bB,\bC\}~~~~~~}\hspace{-0.75cm}&
\frac{1}{2}\|\hspace{-.05cm}\left(\ZZ-\XX\right)\hspace{-.05cm}\hadprod\DD\|_F^2+\hspace{-.05cm}\frac{\mu}{2}\hspace{-.05cm}\left(\|\bA\|^2_F+\hspace{-.05cm} \|\mathbf
B\|_F^2+\hspace{-.05cm}\|\mathbf C\|_F^2\right) \nonumber\\
 &\hspace{-0.8cm}{\rm s. \: to\;}~~\mathbf X_p=\mathbf A \Diag\left[\bbept \bC\right]\mathbf B,~~ \ p=1,\ldots,P
\end{align}
where %$\mathcal T:=\{\XX\mathbb R^{M\times N \times P},\bA\mathbb R^{M\times R},
%\bB\mathbb R^{N\times R},\bC\in \mathbb R^{ P\times R}\}$.
the Frobenius norm of a tensor
is defined as $\|\XX\|_F^2:=\sum_{m}\sum_{n}\sum_{p} \XX^2(m,n,p)$, and the Hadamard product as
$(\XX\hadprod\DD)(m,n,p):=\XX(m,n,p)\DD(m,n,p)$.

Different from the matrix case, it is unclear whether
the  regularization in \eqref{tensor_approximation}
bears any relation with the tensor rank.
Interestingly, the following   analysis  corroborates the capability of
\eqref{tensor_approximation} to produce a low-rank tensor $\bhatZ$, for sufficiently
large $\mu$. In this direction, consider an alternative  completion problem stated in terms
of the normalized tensor representation \eqref{normalized_outerproduct}
\begin{align}\label{atomic norm_23}
\nonumber\bhatZ':=\arg\hspace{-0.9cm}\min_{\{\XX,\bm\gamma,\{\mathbf{u}_r\},\{\mathbf{v}_r\},\{\mathbf{w}_r\}\}}&
\frac{1}{2}\|\left(\ZZ-\XX\right)\hadprod  \DD\|_F^2+\frac{\mu}{2}\|\bm\gamma\|^{2/3}_{2/3} \\
&\hspace{-1.6cm}{\rm s. \: to\;}\mathbf\XX=\sum_{r=1}^R \gamma_r (\mathbf u_r\circ\mathbf v_r\circ\mathbf w_r)
\end{align}
where $\bm\gamma:=[\gamma_1,\ldots,\gamma_R]^T$; the nonconvex $\ell_{2/3}$
(pseudo)-norm is given by $\|\bm{\gamma}\|_{2/3}:=(\sum_{r=1}^R|\gamma_r|^{2/3})^{3/2}$;
and the unit-norm constraint on the factors' columns is left implicit.
Problems \eqref{tensor_approximation} and \eqref{atomic norm_23} are equivalent as established by the following proposition
(its proof is provided in the Appendix.)

\begin{proposition}\label{proposition_2}
The solutions of \eqref{tensor_approximation} and \eqref{atomic norm_23} coincide, i.e.,
$\bhatZ'=\bhatZ$, with optimal
factors related by $\mathbf{\hat a}_r=\sqrt[3]{\hat\gamma_r}\mathbf{\hat u}_r$,   $\mathbf{\hat b}_r=\sqrt[3]{\hat\gamma_r}\mathbf{\hat v}_r$,  and $\mathbf{\hat c}_r=\sqrt[3]{\hat\gamma_r}\mathbf{\hat w}_r$, $r=1,\ldots,R$.
\end{proposition}

\begin{figure}
  \centering
  \centerline{\epsfig{figure=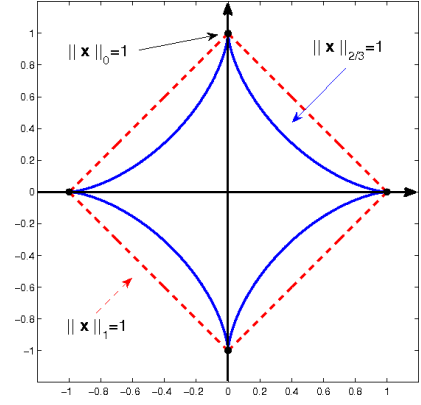,width=0.6\linewidth}}
%  \vspace{1.5cm}
%  \centerline{(c) Result 4}\medskip
\caption{The $\ell_{2/3}$-norm ball compared to the $\ell_{0}$- and $\ell_{1}$-norm balls }\vspace{-.3cm}
\label{fig:balls}
\end{figure}

To further stress the capability of \eqref{tensor_approximation} to produce a low-rank
approximant tensor $\XX$, consider transforming  \eqref{atomic norm_23} once more by
rewriting it in the constrained-error form
\begin{align}\label{atomic norm_23_constrained}
\bhatZ'':=\arg\hspace{-0.7cm}\min_{\{\XX,\bm\gamma,\{\mathbf{u}_r\},\{\mathbf{v}_r\},\{\mathbf{w}_r\}\}}&
\|\bm\gamma\|_{2/3} \\
\nonumber&\hspace{-3.3cm}{\rm s. \: to\;}||\left(\ZZ-\XX\right)\hadprod
\DD||_F^2\leq\sigma^2,\quad\mathbf\XX=\sum_{r=1}^R \gamma_r (\mathbf u_r\circ\mathbf v_r\circ\mathbf w_r).\nonumber
\end{align}
For any value of $\sigma^2$ there exists a corresponding
Lagrange multiplier $\lambda$ such that  \eqref{atomic norm_23} and \eqref{atomic norm_23_constrained}
yield the same solution, under the identity $\mu=2/\lambda$. [Since $f(x)=x^{2/3}$ is an
increasing function, the exponent of $\|\bm\gamma\|_{2/3}$ can be safely  eliminated without affecting
the minimizer of $\eqref{atomic norm_23_constrained}$.] The   $\ell_{2/3}$-norm
$\|\bm\gamma\|_{2/3}$ in \eqref{atomic norm_23_constrained} produces a sparse vector
$\bm \gamma$ when minimized \cite{C07},  sharing this well-documented property of the
$\ell_1$-norm as their norm-one balls, depicted in Fig. \ref{fig:balls}, share the ``pointy geometry''
which is responsible for inducing sparsity.

With \eqref{tensor_approximation} equivalently rewritten as in  \eqref{atomic norm_23_constrained},
its low-rank inducing property is now revealed.
As  $\bm \gamma$ in \eqref{atomic norm_23_constrained} becomes sparse, some of its entries $\gamma_r$
are zeroed, and  the corresponding outer-products $\gamma_r(\mathbf a_r\circ \mathbf b_r\circ\mathbf c_r)$
drop from the sum in \eqref{normalized_outerproduct},  thus lowering the rank of $\XX$.

The next property is a direct consequence of the low-rank promoting property of \eqref{tensor_approximation} as established in Proposition \ref{proposition_2}.

\begin{corollary}
If $\bhatZ$ denotes the solution to problem  \eqref{tensor_approximation} \label{corollary:mu_max},
and $\mu\geq \mu_{\max}:= \|\DD\hadprod\ZZ\|_F^{4/3}$, then $\bhatZ=\mathbf{0}_{M\times N\times P}$.
\end{corollary}

Corollary \ref{corollary:mu_max}  asserts that if the penalty parameter
is chosen large enough, the rank is reduced to the extreme
case $\textrm{rank}(\bhatZ)=\mathbf 0$. To see why this is a non-trivial property,
it is prudent to think of ridge-regression estimates
where similar quadratic regularizers are adopted,
but an analogous property does not hold. In ridge regression one needs to let $\mu\to\infty$ in order to obtain
an all-zero solution. Characterization of $\mu_{\max}$ is also of practical relevance as it provides a frame of reference for tuning the regularization parameter.

Using \eqref{atomic norm_23_constrained}, it is also possible
to relate \eqref{tensor_approximation} with the atomic norm
in \cite{CRPW12}.  Indeed,  the infimum $\ell_1$-norm of
$\bm\gamma$  is a proper norm for $\XX$, named atomic norm,
and denoted by $\|\XX\|_{\mathcal A}:=\|\bm\gamma\|_1$ \cite{CRPW12}.
Thus, by replacing $\|\bm\gamma\|_{2/3}$ with $\|\XX\|_{\mathcal A}$,
\eqref{atomic norm_23_constrained} becomes convex in $\XX$. Still,
the complexity of solving such a variant of \eqref{atomic norm_23_constrained}
resides in that $\|\XX\|_{\mathcal A}$ is generally intractable
to compute \cite{CRPW12}. In this regard, it is remarkable that
arriving to \eqref{atomic norm_23_constrained} had the sole
purpose of demonstrating the low-rank inducing property,
and that \eqref{tensor_approximation} is to be solved by
the algorithm developed in the ensuing section. Such an
algorithm will neither require computing the atomic norm
or PARAFAC decomposition of  $\XX$, nor knowing its rank.
The number of columns in $\bA$, $\bB$, and $\bC$ can be
set to an overestimate of the rank of $\ZZ$, such as the upper bound
$\bar R:=\min\{MN,NP,PM\}\geq \textrm{rank}(\ZZ)$, and the
low-rank of $\XX$ will be induced by regularization as argued
earlier. To carry out a fair comparison, only convergence to
a stationary point of \eqref{tensor_approximation}
will be guaranteed in this paper.

\noindent\textbf{Remark 1:} These insights foster future research
directions for the design of a convex regularizer of the tensor rank.
Specifically, substituting $\rho(\bA,\bB,\bC):=\sum_{r=1}^R(\|\mathbf a_r\|^{3}
+\|\mathbf b_r\|^{3}+\|\mathbf c_r\|^{3})$ for the regularization
term in \eqref{tensor_approximation}, turns $\|\bm\gamma\|_{2/3}$ into
$\|\bm\gamma\|_1=\|\XX\|_{\mathcal A}$ in the equivalent \eqref{atomic norm_23_constrained}.
It is envisioned that with such a modification in place, the acquired
convexity of \eqref{atomic norm_23_constrained} would enable a
reformulation of Proposition \ref{prop:prop_matrix}, providing
conditions for global optimality of the stationary points of
\eqref{tensor_approximation}.

Still, a limitation of \eqref{tensor_approximation} is that it does not allow for
incorporating side information that could be available
in addition to the given  entries $\DD\hadprod\ZZ$.

\noindent\textbf{Remark 2:} In the context of recommender systems, a description
of the users and/or  products through attributes
(e.g., gender, age) or measures of similarity, is typically available. It is
thus meaningful to exploit both known preferences and descriptions
to model the preferences of users~\cite{ABEV09}.
In three-way (samples, genes, conditions) microarray data analysis,
the relative position of single-nucleotide polymorphisms in the
DNA molecule implies degrees of correlation among genotypes~\cite{biocorr}.
These correlations could be available either through a prescribed model, or,
through estimates obtained using a reference tensor $\check \ZZ$.
A probabilistic approach to tensor completion capable
of incorporating such types of extra information is the  subject of the ensuing section.

\section{Bayesian low-rank tensor approximation}\label{sec:Bayesian}

\subsection{Bayesian PARAFAC model}

A probabilistic approach is developed in this section in order to integrate the available statistical information into the tensor imputation setup. To this end, suppose that  the observation noise  is zero-mean, white, Gaussian;  that is
\begin{equation}\label{additive_model} Z_{mnp}=X_{mnp} + e_{mnp};\textrm{ such that } e_{mnp}\sim \mathcal N(0,\sigma^2),\ \ i.i.d..
\end{equation}
Since vectors $\mathbf a_r$ in  \eqref{normalized_outerproduct} are interchangeable, identical distributions are assigned across $r=1,\ldots,R,$ and they are modeled as independent from each other, zero-mean  Gaussian distributed with covariance matrix $\bR_A \in {\mathbb R^{M\times M}}$.
Similarly,  vectors $\mathbf b_r$ and $\mathbf c_r$ are uncorrelated and zero-mean, Gaussian, with covariance matrix $\bR_B$ and $\bR_C$, respectively.
In addition $\mathbf a_r$, $\mathbf b_r$, and  $\mathbf c_r$ are assumed mutually uncorrelated.
And since scale ambiguity is inherently present in the PARAFAC model,   vectors  $\mathbf a_r$, $\mathbf b_r$, and $\mathbf c_r$  are set to have equal power; that is,
\begin{equation}\label{equal_power}
\theta:=\trace(\bR_A)=\trace(\bR_B)=\trace(\bR_C).
\end{equation}

Under these assumptions, the negative of the posterior distribution can be readily written as $\exp(-L(\XX))$, with
\begin{align*}
L(\XX)&=\frac{1}{2\sigma^2}\|(\ZZ-\XX)\hadprod\DD\|_F^2\\
 &+ \frac{1}{2}\sum_{r=1}^R\left(\mathbf a_r^T\bR_A^{-1} \mathbf a_r + \mathbf  b_r^T\bR_B^{-1} \mathbf b_r+ c_r^T\bR_C^{-1} \mathbf c_r\right)\\
 &=\frac{1}{2\sigma^2}\|(\ZZ-\XX)\hadprod\DD\|_F^2+\frac{1}{2}\left[ \trace\left(\mathbf A^T\bR_A^{-1} \mathbf A\right)\right.\\
& \left.+\trace\left(\mathbf B^T\bR_B^{-1} \mathbf B\right)+\trace\left(\mathbf C^T\bR_C^{-1} \mathbf C\right)\right].
\end{align*}
Correspondingly, the MAP estimator of $\XX$ is
\begin{align}
\nonumber \bhatZ:=\arg\hspace{-0.4cm} \min_{\{\XX,\bA,\bB,\bC\}}&\frac{1}{2\sigma^2}\|(\ZZ-\XX)\hadprod\DD\|_F^2+\frac{1}{2}\left[ \trace\left(\mathbf A^T\bR_A^{-1} \mathbf A\right)\right.\\
& \left.+\trace\left(\mathbf B^T\bR_B^{-1} \mathbf B\right)+\trace\left(\mathbf C^T\bR_C^{-1} \mathbf C\right)\right]\nonumber\\
 &\hspace{-1.5cm}{\rm s. \: to\;}\mathbf
X_p=\mathbf A \Diag\left[\bbept \bC\right]\mathbf B^T, \ p=1,\ldots,P\label{tensor_map}
\end{align}
reducing to \eqref{tensor_approximation} when $\bR_A=\mathbf I_M,$ $\bR_B=\mathbf I_N,$ and $\bR_C=\mathbf I_P.$
 This Bayesian approach  interprets the regularization parameter $\mu$ [cf. \eqref{tensor_approximation}] as the noise variance, which is useful in practice to select $\mu$.
The ensuing section explores the advantages of incorporating prior information to the imputation method.
\subsection{Nonparametric tensor decomposition}
Incorporating the information conveyed by  $\bR_A$, $\bR_B$, and $\bR_C$, together with a practical means of finding these matrices can be facilitated by interpreting \eqref{tensor_map} in the context of RKHS \cite{W90}. In particular, the analysis presented next will use the Representer Theorem, interpreted as an instrument for finding the best interpolating function in a Hilbert space spanned by kernels, just as interpolation with sinc-kernels is carried out in the space of bandlimited functions for the purpose of reconstructing a signal from its samples \cite{NS12}.

In this context, it is instructive to look at a tensor $f:\caliM\times\caliN\times\caliP \to \mR$
as a function of three variables $m,n,$ and $p$, living in measurable
spaces $\caliM, \caliN,$ and $\caliP$, respectively.
%
% Identifying  $m,n,$
%and $t$ with the row, column and fiber indexes $m,n,$ and $p$ used so far,
%renders function $f(m,n,p)$ recognizable as a three-way array, but with $x,\ y,$
%and $t$  expanded to incorporate the informative attributes. In the microarray data example, index $n$ may be expanded to $y_n=(\textrm{gene id},\textrm{gene position}).$
%
Generalizing \eqref{tensor_approximation} to this
nonparametric framework, low-rank functions $f$ are formally defined to belong to  the following family
\begin{align*}
\vspace*{-0.25cm}
\nonumber\caliF_R:=&\Large\{f:\caliM\hspace{-.08cm}\times\hspace{-.08cm}\caliN\hspace{-.08cm}\times\hspace{-.08cm}\caliP\hspace{-.1cm}\rightarrow\hspace{-.1cm}\mathbb R:\ f(m,n,p)=\hspace{-.08cm}\sum_{r=1}^R a_r(m) b_{r}(n) c_{r}(p)\\
 & \textrm{ such that } a_{r}(m)\in
\caliHm,\ b_{r}(n)\in\caliHn , \  c_{r}(p)\in\caliHp\Large\}
\end{align*}
%
%low-rank functions
%$f(x,y,t)=\sum_{r=1}^{R}a_r(x)b_r(y)c_r(t)$, and estimate them
%for preselected kernels $k_\caliM$, $k_\caliN$ and $k_\caliN$;   see also
%\cite{bmg11tsp,bmg11icassp,ssm97,low_rank_kernels}.
%
%
%\noindent {\bf Technical approach.} We have recently developed an
%interpolation method to fit low-rank functions to the available
%data, while incorporating available attributes.
%%
where $\caliHm$, $\caliHn$, and $\caliHp$ are Hilbert spaces constructed from
specified kernels $k_\caliM$, $k_\caliN$ and $k_\caliP$, defined over $\caliM$, $\caliN$, and
$\caliP$,
 while $R$
is an initial overestimate of the rank of $f$.

 The following
nonparametric fitting criterion is adopted for finding the best   $\hat f_R$ interpolating data $\{z_{mnp}:\ \delta_{mnp}=1\}$
\begin{align}
\vspace*{-0.25cm}
\nonumber\hat f_R:=\arg&\min_{f\in \caliF_R}
\summ\sumn\sump \delta_{mnp}(z_{mnp}-f(m,n,p))^2\\
&+\frac{\mu}{2} \sumr
\left(\|a_{r}\|^2_{\caliHm}+\|b_{r}\|^2_{\caliHn}+\|c_{r}\|^2_{\caliHp}\right) \;.
\label{kernel_estimator}
\end{align}
It is shown in the Appendix that leveraging the Representer Theorem, the minimizer of \eqref{kernel_estimator} admits a finite dimensional representation in terms of $k_\caliM$, $k_\caliN$ and $k_\caliP$,
\begin{align}
\hat f_R(m,n,p){}={}\bk_{\caliM}^T(m)\bK_{\caliM}^{-1} \bA \Diag\left[\bk_{\caliP}^T(p)\bK_{\caliP}^{-1}\bC\right]\bB^T
\bK_{\caliN}^{-1}\bk_{\caliN}(n)\label{fR_kernel}
\end{align}
where vector $\mathbf k_{\caliM}(m)$ and matrix $\bK_{\caliM}$ have entries $k_{\caliM}(m,m')$, $m,m'=1,\ldots,M$;  and  where $\mathbf k_{\caliN}(n)$, $\bK_{\caliN}$, $\mathbf k_{\caliP}(p)$, and  $\bK_{\caliP}$ are correspondingly defined  in terms of  $k_\caliN$ and $k_\caliP$.
%Then the function $f$ in \eqref{f_quadratic_form} is evaluated in
%each of the training points and the results ordered in matrix $\bf
%F$
It is also shown in the Appendix that the coefficient matrices $\bA$, $\bB$, and $\bC$ can be found
by solving
%\begin{equation}
%\bf F:=((f(x_n,y_m))_{N\times M}= \bK_{\caliM}^T \bA
%\bK_{\caliN}
%\end{equation}
%
%
\begin{align}
&\nonumber\min_{\bA,\bB,\bC}
%\XX\in \mathbb^{M\times N\times P}   \bA\in\mathbb
%R^{M\times R}\\  \bB\in\mathbb R^{N\times R} \bC\in\mathbb R^{P\times R}}  }
%\|(\ZZ-
%\XX)\hadprod \DD\|_F^2\\
\sump \left\|\left(\bZ_p- \bA \Diag\left[\bbept \bC\right]\bB^T\right)\hadprod \bDelta_p\right\|_F^2\\
\nonumber&\hspace{-0.04cm}+\hspace{-0.04cm}\frac{\mu}{2}\hspace{-0.04cm}
\left(\trace(\hspace{-0.04cm}\bA^T\hspace{-0.04cm}\bK_{\caliM}^{-1}
\hspace{-0.04cm}\bA\hspace{-0.04cm})\hspace{-0.04cm}+\hspace{-0.04cm}\trace(\hspace{-0.04cm}\bB^T\hspace{-0.04cm}\bK_{\caliN}^{-1} \hspace{-0.04cm}\bB\hspace{-0.04cm})\hspace{-0.04cm}+\hspace{-0.04cm}\trace(\hspace{-0.04cm}\bC^T\hspace{-0.04cm}\bK_{\caliP}^{-1}\hspace{-0.04cm} \bC\hspace{-0.04cm})\right)\\
%&\textrm{s.\ to}\ \bX_p=\bK_{\caliM} \bA \bD_{(\bK_{\caliP}\bC),p}\bB^T\bK_{\caliN}\\
&\hspace{1cm}\hspace{-0.04cm}\textrm{s.\ to}\  \bA\in\mathbb R^{M\times R},\
\bB\in\mathbb R^{N\times R},\ \bC\in\mathbb R^{P\times R}\label{Kernel_tensor approximation}.
\end{align}

Problem \eqref{Kernel_tensor approximation} reduces to \eqref{tensor_approximation} when the
side information is discarded by selecting  $k_{\caliM}$, $k_{\caliN}$ and $k_{\caliP}$
as Kronecker deltas, in which case $\bK_{\caliM}$, $\bK_{\caliN}$, and $\bK_{\caliP}$ are identity matrices.  In the general case, \eqref{Kernel_tensor approximation} yields the
sought nonlinear low-rank approximation method for $f(m,n,p)$ when
combined with \eqref{fR_kernel}, evidencing the equivalence between \eqref{kernel_estimator} and \eqref{tensor_map}.

Interpreting \eqref{kernel_estimator}  as an interpolator renders \eqref{tensor_map} a natural  choice  for tensor imputation, where in general, missing entries are to be inserted by connecting them to surrounding points on the three-dimensional arrangement.
Relative to \eqref{tensor_approximation}, this RKHS perspective also highlights \eqref{tensor_map}'s extra smoothing and extrapolation capabilities. Indeed, by capitalizing on the similarities captured by $\bK_{\caliM}$, $\bK_{\caliN}$ and $\bK_{\caliP}$, \eqref{Kernel_tensor approximation} can recover completely missing slices. This  feature is not shared by imputation methods that leverage  low-rank
 only, since these require at least one point in the slice to build on colinearities.    Extrapolation is also  possible in this sense. If for instance $\bK_{\caliM}$ can be expanded to capture a further point $M+1$ not in the original set, then a new slice of data can be predicted by \eqref{fR_kernel} based on its correlation $k_{\caliM}(M+1)$ with the available entries.
These extra capabilities will be exploited in Section \ref{sec:numerical_tests}, where correlations are leveraged for the imputation of MRI data. The  method described by \eqref{tensor_map} and \eqref{Kernel_tensor approximation} can be applied to matrix completion by just setting entries of $\bC$ to one, and  can be extended to higher-order dimensions with a straightforward alteration of the algorithms and theorems throughout this paper.  %In addition, \eqref{kernel_estimator} reduces to \eqref{tensor_approximation} when the kernels are selected as Kronecker deltas

 Identification of covariance matrices $\bR_A$, $\bR_B$, and $\bR_C$ with  kernel matrices  $\bK_{\caliM}$, $\bK_{\caliN}$ and $\bK_{\caliP}$ is the  remaining aspect to clarify in the connection between  \eqref{tensor_map} and \eqref{Kernel_tensor approximation}. It is apparent from \eqref{tensor_map} and \eqref{Kernel_tensor approximation} that correlations between columns of the factors are reflected in similarities between the tensor slices, giving rise to the opportunity of obtaining one from the other. This aspect is explored next.
\subsection{Covariance estimation}

To implement \eqref{tensor_map}, matrices $\bR_A$, $\bR_B$, and $\bR_C$ must be postulated a priori,
or alternatively replaced by their sample estimates. Such estimates  need a training set of vectors $\mathbf a$, $\mathbf b$,
and $\mathbf c$ abiding to the Bayesian model just described, and this requires
PARAFAC decomposition of training data. In order to abridge this procedure,
it is convenient to inspect how  $\bR_A$, $\bR_B$, and $\bR_C$ are related to
their kernel counterparts.

Based on the equivalence between the standard RKHS interpolator and the linear mean-square error estimator \cite{RW06}, it is useful to re-visit  the  probabilistic framework and identify kernel similarities between slices of $\XX$ with their mutual covariances.
Focusing on the tube  dimension of $\XX$, one can write
 $\bK_{\caliP}(p',p):=\E(\trace(\bX_{p'}^T\bX_{p}))$, that is, the covariance
between slices $\bX_{p'}$ and $\bX_{p}$  taking $\langle\bX,\mathbf Y\rangle:= \trace(\bX^T\mathbf Y)$
as the standard inner product in the matrix space.
Under this alternative definition for $\bK_{\caliP}$, and corresponding definitions for  $\bK_{\caliN}$, and $\bK_{\caliM}$, it is shown in the Appendix that
\begin{align}
\bK_{\caliM}=\theta^2 \bR_A,\quad \bK_{\caliN}=\theta^2 \bR_B, \quad \bK_{\caliP}=\theta^2 \bR_C\label{KthetaR}
\end{align}
 and that  $\theta$ is related to the second-order moment of $\XX$ by
\begin{align}
\E\|\XX\|_F^2=R\theta^3\label{ERtheta3}.
\end{align}

Since sample estimates for $\bK_{\caliM}$, $\bK_{\caliN}$, $\bK_{\caliP}$, and $\E\|\XX\|_F$ can be readily obtained from the tensor data, \eqref{KthetaR} and \eqref{ERtheta3}   provide an agile means of estimating $\bR_A$, $\bR_B$, and $\bR_C$ without
requiring PARAFAC decompositions over the set of training tensors.

This strategy remains valid when kernels are not estimated from data.
One such  case emerges in  collaborative filtering of user preferences \cite{ABEV09}, where the similarity of two users is modeled as a function of attributes; such age or  income.

\subsection{Block successive upper-bound minimization algorithm}

An iterative algorithm is developed here for solving \eqref{tensor_map}, by cyclically minimizing the cost over  $\bA$, $\bB$, and $\bC$. In the first  step of the cycle the cost in \eqref{tensor_map} is minimized with respect to (w.r.t.) $\bA$ considering $\bB$ and $\bC$ as parameters. Accordingly, the partial cost to minimize reduces to
\begin{align}
f(\bA):=\frac{1}{2}\|\left(\ZZ-\XX\right)\hadprod  \DD\|_F^2+\frac{\mu}{2}\trace\left(\mathbf A^T\bR_A^{-1} \mathbf A\right)\label{f_ls}
\end{align}
where $\mu$ was identified with and substituted for $\sigma^2$.
Function \eqref{f_ls} is quadratic in $\bA$ and can be readily minimized after re-writing it in terms of $\mathbf a:=\textrm{vec}(\bA)$
[see \eqref{flsa_appendix} in the  Appendix]. However, such an approach becomes computationally infeasible for other than small datasets, since it involves  storing  $P$ matrices of dimensions $NM\times MR$, and solving  a linear system of $MR\times MR$ equations. The alternative pursued here to overcome this obstacle relies on the so-called block successive upper-bound minimization (BSUM) algorithm \cite{luo_bsum}.

In BSUM one minimizes a judiciously chosen upper-bound $g(\bA,\bar\bA)$ of $f(\bA)$, which: i) depends
on the current iterate $\bar{\bA}$; ii) should
be simpler to optimize; and iii) satisfies certain local-tightness  conditions;
see also~\cite{luo_bsum} and properties i)-iii) below.

For $\bar\bA$ given, consider the  function
\begin{align}
g(\bA,\bar\bA)&:=\frac{1}{2}\|\left(\ZZ-\XX\right)\hadprod  \DD\|_F^2 \label{g_ls}\\
&+{\mu}\left(\frac{\lambda}{2} \trace\left(\mathbf A^T \mathbf A\right) -\trace(\bm\Theta^T\bA)+\frac{1}{2}\trace(\bm\Theta^T\bar\bA)\right)\nonumber
\end{align}
where $\lambda:=\lambda_{\max}(\bR_A^{-1})$ is the  maximum eigenvalue of  $\bR_A^{-1}$, and  $\bm\Theta:=\lambda\mathbf I -\bR_A^{-1}$.
The following properties of $g(\bA,\bar\bA)$ imply that it majorizes $f(\bA)$ at $\bar\bA$,
satisfying the technical conditions required for the convergence of BSUM (properties i)-iii) are established in the the proof of Lemma \ref{lemma:majorizing} in the Appendix).
\begin{itemize}
\item[i)]$f(\bar\bA)=g(\bar\bA,\bar\bA)$;
\item[ii)]$\frac{d}{d\bA}f(\bA)|_{\bA=\bar\bA}=\frac{d}{d\bA}g(\bA,\bar\bA)|_{\bA=\bar\bA}$; and,
\item[iii)]$f(\bA)\leq g(\bA,\bar\bA),~\forall \bA$.
\end{itemize}

The computational advantage of minimizing $g(\bA,\bar\bA)$  in place of  $f(\bA)$ comes from $g(\bA,\bar\bA)$ being separable across rows of
$\bA$. To see this, consider the Kathri-Rao product  $\bm\Pi:=\bC\krprod \bB:=[\mathbf c_1\otimes\mathbf b_1,\ldots \mathbf c_{R}\otimes\mathbf b_{R}]$, defined by the column-wise Kronecker products $\mathbf c_r\otimes \mathbf b_r$. Let also  matrix $\bZ:=[\bZ_1,\ldots,\bZ_P]
\in\mathbb{N}^{M\times NP}$ denote the  unfolding of $\ZZ$ along its tube dimension, and likewise
for $\bDelta:=[\bDelta_1,\ldots,\bDelta_P]\in\{0,1\}^{M\times NP}$ and  $\bX:=[\bX_1,\ldots,\bX_P]\in\mathbb{R}_+^{M\times NP}$.
Then, using the following identity~\cite{kolda_poisson}
\begin{align}\label{krprod}
\bX:=[\bX_1,\ldots,\bX_P]=\bA \bm\Pi^T.
\end{align}
it is possible to rewrite \eqref{g_ls} as
\begin{align*}
 g(\bA,\bar\bA)&:=\frac{1}{2}\|\left(\bZ-\bA \bm\Pi^T\right)\hadprod  \bDelta\|_F^2\\
&\nonumber+{\mu}\left(\frac{\lambda}{2} \trace\left(\mathbf A^T \mathbf A\right) -\trace(\bm\Theta^T\bA)+\frac{1}{2}\trace(\bm\Theta^T\bar\bA)\right)
\end{align*}
which can be decomposed as
%$g_m(\mathbf a_m,\mathbf{\bar a}_m ):=\frac{1}{2}\|\left(\mathbf z_m-\bm\Pi \mathbf a_m \right)\hadprod  \bm \delta_m \|_2^2
%+ \frac{\mu}{2} \left(\lambda \|\mathbf a_m\|^2 + \bm \theta_m^T\mathbf a_m + \bm\theta_m^T\mathbf{\bar a_m}\right)$\
%
\begin{align}
\nonumber g(\bA,\bar\bA)& = \summ  \left[\frac{1}{2}\| \bm \delta_m \hadprod\mathbf z_m-\Diag( \bm \delta_m)\bm\Pi \mathbf a_m  \|_2^2\right.\\
&\left.+ \frac{\mu}{2} \left(\lambda \|\mathbf a_m\|^2 + \bm \theta_m^T\mathbf a_m + \bm\theta_m^T\mathbf{\bar a}_m\right)\right]\label{g_lsstandars}
\end{align}
where  $\mathbf z_m^T$, $\mathbf a_m^T$, $\bm \delta_m^T$,
$\bm \theta_m^T$, and $\mathbf{\bar a_m}^T$, represent the $m$-th rows of matrices  $\bZ$, $\bA$, $\bDelta$, $ \bm\Theta$, and  $\bar\bA$, respectively.
Not only  \eqref{g_lsstandars}  evidences the separability of \eqref{g_ls} across rows of $\bA$, but it also  presents each of its summands in a  standardized quadratic form that can be readily minimized by equating its gradient to zero. Accordingly, the majorization strategy reduces the computational load to $R$ systems of $M$ equations that can be solved in parallel.
Collecting the solution of such quadratic programs into the rows of a matrix $\bA^*$  yields the minimizer of \eqref{g_ls},  and the update $\bA\leftarrow\bA^*$ for the BSUM cycle. Such a procedure is presented in Algorithm \ref{table-LRTI}, where analogous updates for $\bB$ and $\bC$ are carried out cyclically.

%\algsetup{indent=2em}
 \begin{algorithm}[t]
% \algblock[Update]{Update}{End}
% \algblock[function]{function}{end_function}
 \caption{: Low-rank tensor imputation (LRTI)} \small{
 %\algblock[UPDATE]{Start}{End}
 \begin{algorithmic}[1]
\Function{update\_factor}{$\bA,\bR,\bm\Pi,\DD,\ZZ,\mu$}
             \State Set $\lambda=\lambda_{\max}(\bR^{-1})$
             \State Unfold $\DD$ and  $\ZZ$ over dimension of $\bbA$ into $\bDelta$ and  $\bZ$
             \State Set $\bm\Theta=(\lambda \mathbf I-\bR^{-1})\bA$
             \For{$m=1,\ldots,M$}
                          \State Select rows $\mathbf z_m^T$,  $\bm \delta_m^T$, and $\bm \theta_m^T$, and set $\bD_m= \Diag( \bm \delta_m)$
                          \State Compute $\mathbf a_m=(\bm\Pi^T \bD_m \bm \Pi +\lambda\mu\mathbf I)^{-1}(\bm\Pi^T\bD_m \mathbf z_m+\mu \bm \theta_m)$
                          \State Update $\bA$ with row $\mathbf a_m^T$
                             \EndFor

\State             \Return $\bA$
  \EndFunction

 \State Initialize $\bA$, $\bB$ and $\bC$ randomly.
      \While {$|\textrm{cost}-\textrm{cost\_old}|<\epsilon$ }
%\State Update $\bA=UPDATE_FACTOR$%(\bA,\bR_A,(\bB\odot \bC),\DD,\ZZ,\mu)$
\State  $\bA=$ \Call{update\_factor}{$\bA,\bR_A,(\bC\odot \bB),\DD,\ZZ,\mu$}
\State  $\bB=$ \Call{update\_factor}{$\bB,\bR_B,(\bA\odot \bC),\DD,\ZZ,\mu$}
\State  $\bC=$ \Call{update\_factor}{$\bC,\bR_C,(\bB\odot \bA),\DD,\ZZ,\mu$}
  \State   Recalculate cost in $\eqref{tensor_map}$
         \EndWhile\\
 \Return  $\XX$ with slices $\mathbf {\hat X_p}=\bA\Diag(\mathbf e_p^T\bC)\bB^T$
 \end{algorithmic}}
 \label{table-LRTI}
 \end{algorithm}
By virtue of properties i)-iii) in Lemma \ref{lemma:majorizing}, convergence of Algorithm \ref{table-LRTI} follows readily
from that of the BSUM algorithm~\cite{luo_bsum}.

\begin{proposition}\label{prop:convergenceLRTI}
The iterates for $\bA$, $\bB$ and $\bC$ generated by Algorithm \ref{table-LRTI} converge to a stationary point of
\eqref{tensor_map}.
\end{proposition}

\section{Inference for low-rank Poisson tensors}\label{sec:poisson}

Adoption of the LS criterion in \eqref{tensor_approximation}  assumes in a Bayesian setting
that the random  $\ZZ$ is Gaussian distributed.
This section deals with a Poisson-distributed tensor $\ZZ$, a natural alternative to the Gaussian
model when integer-valued data are obtained by counting independent events~\cite{kolda_poisson}.
Suppose that the entries $z_{mnp}$ of $\ZZ$ are Poisson distributed, with probability mass function
\begin{align}\label{poisson_model}
P(z_{mnp}=k)=\frac{x_{mnp}^k e^{-x_{mnp}}}{k!}
\end{align}
and means given by the corresponding entries in tensor $\XX$.
For mutually-independent $\{z_{mnp}\}$, the log-likelihood $l_{\DD}(\ZZ;\XX)$ of  $\XX$
given data $\ZZ$ only on the entries specified by $\DD$, takes the
form
\begin{align}\label{llpoisson}
l_{\DD}(\ZZ;\XX)&=\summ\sumn\sump \delta_{mnp} [z_{mnp}\log(x_{mnp})-x_{mnp}]
\end{align}
after dropping terms $\log(z_{mnp}!)$ that do not depend on $\XX$.

The choice of the Poisson distribution in \eqref{poisson_model} over a
Gaussian one for counting data, prompts minimization of the K-L divergence \eqref{llpoisson}
instead of LS as a more suitable criterion~\cite{kolda_poisson}. Still, the entries of $\XX$ are not coupled in
 \eqref{llpoisson}, and a binding PARAFAC modeling assumption
is natural for feasibility of the tensor approximation task under missing data.
%On top of that, augmenting \eqref{llpoisson} with a Frobenius-norm regularizing term on the model factors
%offers an attractive way of controlling the rank of the solution $\hat{\ZZ}$~\cite{tensors_tsp_12}, and  avoid
%well-known indeterminacies of the PARAFAC model~\cite{kolda_regularize}.
Mimicking the method for Gaussian data, (nonnegative) Gaussian priors are assumed for the factors of the PARAFAC decomposition.
Accordingly, the MAP estimator of $\XX$ given Poisson-distributed
data (entries of $\ZZ$ indexed by $\DD$) becomes
\begin{align}
\nonumber&\bhatZ:=\arg\hspace{-0.7cm} \min_{\{\XX,\bA,\bB,\bC\}\in\mathcal T} \summ\sumn\sump
\delta_{mnp} (x_{mnp}-z_{mnp}\log(x_{mnp})) \\
&\hspace{-.2cm}+\frac{\mu}{2}\left[\trace\left(\mathbf A^T\bR_A^{-1} \mathbf A\right)\hspace{-0.1cm}+\hspace{-0.1cm}
\trace\left(\mathbf B^T\bR_B^{-1} \mathbf B\right)\hspace{-0.1cm}+\hspace{-0.1cm}\trace\left(\mathbf C^T\bR_C^{-1} \mathbf C\right)\right]
\label{poisson_map}
\end{align}
%
%It is apparent from \eqref{poisson_map} that the mutually-independent columns of
%$\mathbf{A}$ are modeled as $\mathbf{a}_r\sim\mathcal{N}(\mathbf{0},\bR_A)$, and likewise for
%$\mathbf{B}$ and $\mathbf{C}$. Relative to \eqref{tensor_approximation}, through these covariances one can
%incorporate side information in the form of correlations among tensor slices (along the row, column, and tube dimensions),
%and enhance the estimation performance when large amounts of data are missing. Interestingly,  this also endows \eqref{poisson_map}
%with \emph{prediction} (or extrapolation) capabilities, useful when full slices of data in
%$\ZZ$ are not observed~\cite{tensors_tsp_12}.
over the feasible set $\mathcal T\hspace{-0.1cm}:=\hspace{-0.1cm}\{\XX,\bA,\bB,\bC:\bA\geq\mathbf0,\bB\geq\mathbf0,\bC\geq\mathbf0,$ $\bX_p=\mathbf A \Diag\left[\bbept \bC\right]\mathbf B^T,~ \ p=1,\ldots,P\}$, where the symbol $\geq$ should be understood to imply entry-wise nonegativity.

With the aid of  Representer's Theorem, it is also possible to  interpret \eqref{poisson_map} as a variational estimator in RKHS, with K-L analogues to  \eqref{kernel_estimator}-\eqref{Kernel_tensor approximation}, so that the conclusions thereby regarding smoothing, prediction  and prior covariance estimation carry over to the low-rank Poisson imputation method \eqref{poisson_map}.
\subsection{Block successive upper-bound minimization algorithm}\label{ssec:bsum}
A K-L counterpart of the LRTI algorithm  is developed in this section, that provably converges to a stationary point
of \eqref{poisson_map}, via  an  alternating-minimization iteration which optimizes \eqref{poisson_map} sequentially w.r.t. one factor matrix, while holding the others fixed.

In the sequel, the goal is to arrive at a suitable expression for the cost  in
\eqref{poisson_map}, when viewed only as a function of e.g., $\bA$. To this end, let matrix $\bZ:=[\bZ_1,\ldots,\bZ_P]
\in\mathbb{N}^{M\times NP}$ denote the unfolding of $\ZZ$ along its tube dimension, and likewise
for $\bDelta:=[\bDelta_1,\ldots,\bDelta_P]\in\{0,1\}^{M\times NP}$ and  $\bX:=[\bX_1,\ldots,\bX_P]\in\mathbb{R}_+^{M\times NP}$.
Based on these definitions, \eqref{llpoisson} can be written as
\begin{align}
l_{\bDelta}(\bZ;\bX)=\mathbf 1_M^T (\bm\Delta\hadprod [\bX -\bZ\hadprod \log(\bX)])\mathbf 1_{NP}\label{ll_poison_matrix}
\end{align}
where  $\mathbf 1_M$, $\mathbf 1_{NP}$  are all-one vectors of dimensions $M$ and $NP$ respectively,
and $\log(\cdot)$ should be understood  entry-wise.
The log-likelihood in \eqref{ll_poison_matrix} can be expressed in terms of
$\bA$, and  the Kathri-Rao product $\bm\Pi:=\bB\krprod \bC$ by resorting again to  \eqref{krprod}.
Substituting \eqref{krprod}  into \eqref{ll_poison_matrix} one arrives at the desired expression for
the cost in \eqref{poisson_map} as a function of $\bA$, namely
 %
%\algsetup{indent=2em}
 \begin{algorithm}[t]
% \algblock[Update]{Update}{End}
% \algblock[function]{function}{end_function}
 \caption{: Low-rank Poisson-tensor imputation (LRPTI)} \small{
 %\algblock[UPDATE]{Start}{End}
 \begin{algorithmic}[1]
\Function{update\_factor}{$\bA,\bR,\bm\Pi,\DD,\ZZ,\mu$}
             \State Set $\lambda=\lambda_{\max}(\bR^{-1})$
             \State Unfold $\DD$ and  $\ZZ$ over dimension of $\bbA$ into $\bDelta$ and  $\bZ$
             \State Compute $\mathbf S=\frac{\bA}{\lambda\mu}\hadprod\left( \frac{\bDelta\hadprod\bZ}{\bA\bm\Pi^T}\bm\Pi \right)$ (element-wise division)
             \State Compute $\mathbf T=\frac{1}{2\lambda\mu}\left(\mu(\lambda\mathbf I-\bR^{-1})\bA -\bDelta\bm\Pi\right)$
             \State Update $\bA$ with entries $a_{mr}=t_{mr}+\sqrt{t_{mr}^2+s_{mr}}$
\State             \Return $\bA$
  \EndFunction

 \State Initialize $\bA$, $\bB$ and $\bC$ randomly.
      \While {$|\textrm{cost}-\textrm{cost\_old}|<\epsilon$ }
%\State Update $\bA=UPDATE_FACTOR$%(\bA,\bR_A,(\bB\odot \bC),\DD,\ZZ,\mu)$
\State  $\bA=$ \Call{update\_factor}{$\bA,\bR_A,(\bC\odot \bB),\DD,\ZZ,\mu$}
\State  $\bB=$ \Call{update\_factor}{$\bB,\bR_B,(\bA\odot \bC),\DD,\ZZ,\mu$}
\State  $\bC=$ \Call{update\_factor}{$\bC,\bR_C,(\bB\odot \bA),\DD,\ZZ,\mu$}
  \State   Recalculate cost in $\eqref{poisson_map}$
         \EndWhile\\
 \Return  $\XX$ with slices $\mathbf {\hat X_p}=\bA\Diag(\mathbf e_p^T\bC)\bB^T$
 \end{algorithmic}}
 \label{table-LRPTI}
 \end{algorithm}
\begin{align*}
f(\bA):={}& \mathbf 1_M^T (\bm\Delta\hadprod [\bA \bm\Pi -\bZ\hadprod\log(\bA  \bm\Pi^T) ])\mathbf 1_{NP}\nonumber\\
&+\frac{\mu}{2}\trace\left(\mathbf A^T\bR_A^{-1} \mathbf A\right).%\label{poisson_fA}
\end{align*}
A  closed-form minimizer $\bA^\star$ for $f(\bA)$ is not available, but since $f(\bA)$ is convex  one could
in principle resort to an iterative procedure to obtain $\bA^\star$. To avoid extra inner iterations,
the approach here relies again on the BSUM algorithm~\cite{luo_bsum}.

For $\bar \bA$ given, consider the separable function
\begin{align}\label{eq:major_g}
g(\bA,\bar\bA)\hspace{-0.05cm}:=\hspace{-0.05cm}\mu\lambda\hspace{-0.05cm}\sum_{m,r=1}^{M,R}\hspace{-0.05cm} \hspace{-0.05cm}\Big(\frac{a_{mr}^2}{2}-2t_{mr}a_{mr}-s_{mr}\log(a_{mr})+u_{mr}\Big)
\end{align}
where $\lambda:=\lambda_{\max}(\bR_A^{-1})$ is the largest eigenvalue of $\bR_A^{-1}$, and the
parameters $s_{rm}$, $t_{rm}$, and $u_{rm}$ are defined in terms of $\bar{\bA},$ $\bZ$, $\bDelta$, $\bm\Pi,$ and $\bm\Theta:=\left({\lambda}\mathbf I-\bR_A^{-1}\right)\bar{\bA}$ by
\begin{align*}
s_{mr}&:=\frac{1}{\lambda\mu}\sum_{k=1}^{NP}\frac{\delta_{mk}z_{mk}\bar a_{mr}\pi_{kr}}{\sum_{r'=1}^R \bar a_{mr'}\pi_{kr'}},\\
t_{mr}&:=\frac{1}{2\lambda\mu} \Big( \mu\theta_{mr}-\sum_{k=1}^{NP}\pi_{kr}\delta_{mk}\Big)
\end{align*}
and $u_{mr}:=\frac{1}{\lambda\mu}\left(\theta_{mr}\bar a_{mr}+\sum_{k=1}^{NP}\delta_{mk}z_{mk}\bar a_{mr}\pi_{kr} \upsilon_{mrk}\right)$,  with $\upsilon_{mrk}\hspace{-0.05cm}:=\log(\bar a_{mr}\pi_{kr}/\sum_{r'=1}^R \bar a_{mr'}\pi_{kr'})/\sum_{r'=1}^R \bar a_{mr'}\pi_{kr'}.$
As asserted in the following lemma,  $g(\bA,\bar\bA)$ majorizes $f(\bA)$ at $\bar\bA$
and satisfies the technical conditions required for the convergence of BSUM (see the Appendix for a proof.)

\begin{lemma}\label{lemma:majorizing}
Function $g(\bA,\bar\bA)$ satisfies the following properties
\begin{itemize}
\item[i)]$f(\bar\bA)=g(\bar\bA,\bar\bA)$;
\item[ii)]$\frac{d}{d\bA}f(\bA)|_{\bA=\bar\bA}=\frac{d}{d\bA}g(\bA,\bar\bA)|_{\bA=\bar\bA}$; and,
\item[iii)]$f(\bA)\leq g(\bA,\bar\bA),~\forall \bA$.
\end{itemize}
Moreover, $g(\bA,\bar\bA)$ is minimized at $\bA=\bA_g^\star$ with entries
$a_{g,mr}^{\star}:=t_{mr}+\sqrt{t_{mr}^2+s_{mr}}$.
\end{lemma}

\noindent Lemma \ref{lemma:majorizing} highlights the reason behind adopting $g(\bA,\bar\bA)$
in the proposed block-coordinate descent algorithm:  it is separable across the entries of its matrix argument [cf. \eqref{eq:major_g}],
and hence it admits a closed-form minimizer given by the $MR$ scalars $a_{g,mr}^{\star}$.
The updates $\bA\leftarrow \bA^*_g$ are tabulated under Algorithm \ref{table-LRPTI} for solving \eqref{poisson_map}, where analogous updates for $\bB$ and $\bC$ are carried out cyclically.

By virtue of properties i)-iii) in Lemma \ref{lemma:majorizing}, convergence of Algorithm \ref{table-LRPTI} follows readily
from the general convergence theory available for the BSUM algorithm~\cite{luo_bsum}.

\begin{proposition}\label{prop:convergenceLRPTI}
The iterates for $\bA$, $\bB$ and $\bC$ generated by Algorithm \ref{table-LRPTI} converge to a stationary point of
\eqref{poisson_map}.
\end{proposition}

A related algorithm, abbreviated as CP-APR can be found in \cite{kolda_poisson}, where the objective is to find the tensor's low-rank factors per se. The LRPTI algorithm here generalizes CP-APR by focusing on recovering missing data, and incorporating prior information through rank regularization. In terms of convergence to a stationary point, the added regularization allows for lifting the assumption on the linear independence  of the rows of  $\bm\Pi$, as required by  CP-APR \cite{kolda_poisson} -  an assumption without a straightforward validation since iterates $\bm\Pi$ are not  accessible beforehand.

\section{Numerical Tests}\label{sec:numerical_tests}
\subsection{Simulated Gaussian data}
Synthetic tensor-data of dimensions $M\times N\times P=16\times 4\times 4$ were generated according to the Bayesian tensor model described in Section \ref{sec:Bayesian}. Specifically, entries of $\ZZ$ consist of realizations of Gaussian random variables generated according to \eqref{additive_model}, with means specified by entries of $\XX$ and variance scaled to yield an SNR of $-20$dB . Tensor $\XX$  is constructed from factors $\bA$, $\bB$ and $\bC,$ as in \eqref{parafac}. Matrices $\bA,$ $\bB$, and $\bC$ have $R=6$ columns containing realizations of independent zero-mean, unit-variance, Gaussian random variables.

 A quarter of the entries of $\ZZ$ were removed at random and reserved to evaluate performance. The remaining seventy five percent of the data were used to recover $\ZZ$ considering the removed data as missing entries. Method \eqref{tensor_approximation} was employed for recovery, as implemented by the LRTI Algorithm, with   regularization  $\frac{\mu}{2}(\|\bA\|_F^2+\|\bB\|_F^2+\|\bC\|_F^2)$ resulting from setting $\bR_A=\bI_M$, $\bR_B=\bI_N$, and $\bR_C=\bI_P$.

\begin{figure}
  \includegraphics[width=0.8\linewidth]{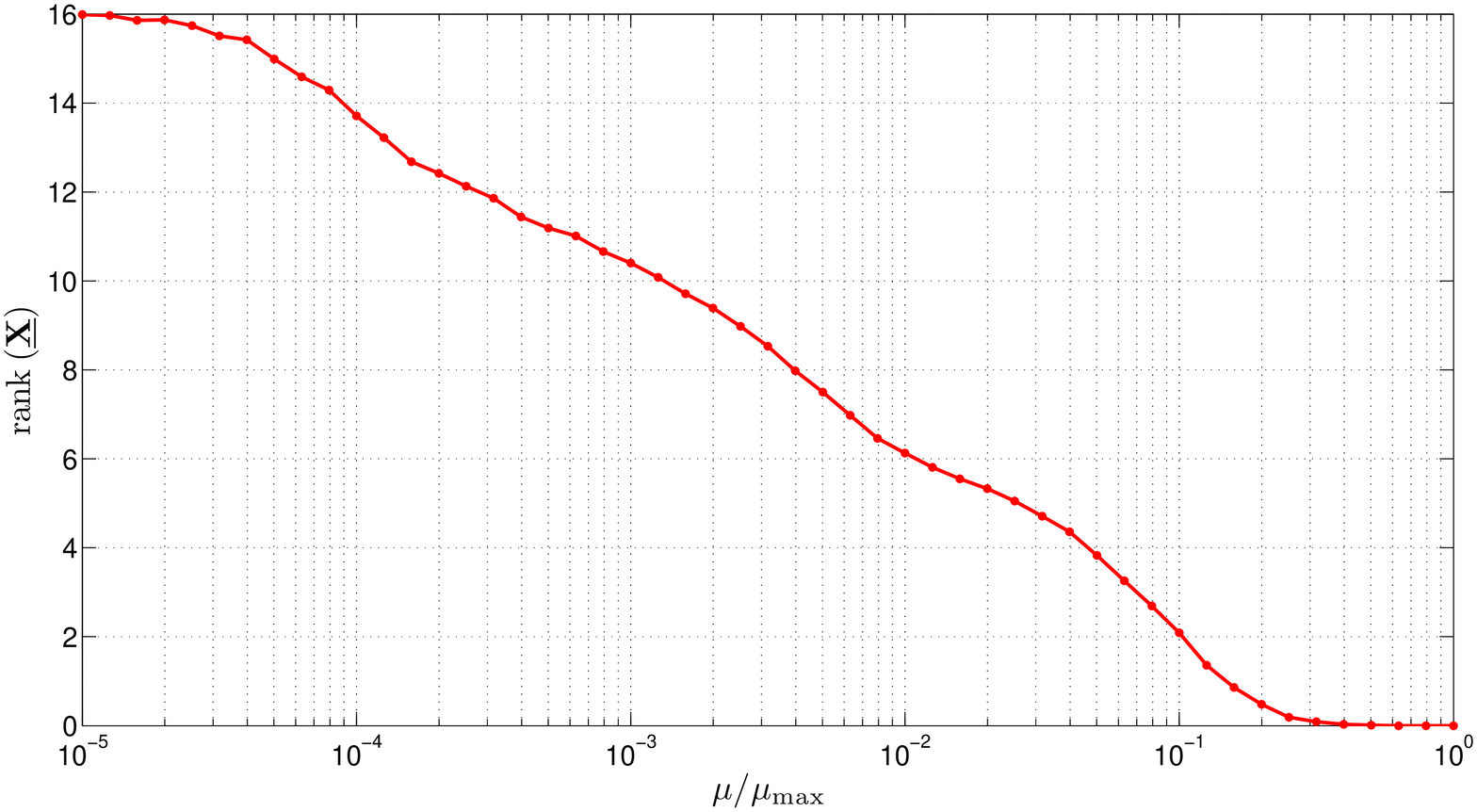}
   \includegraphics[width=0.8\linewidth]{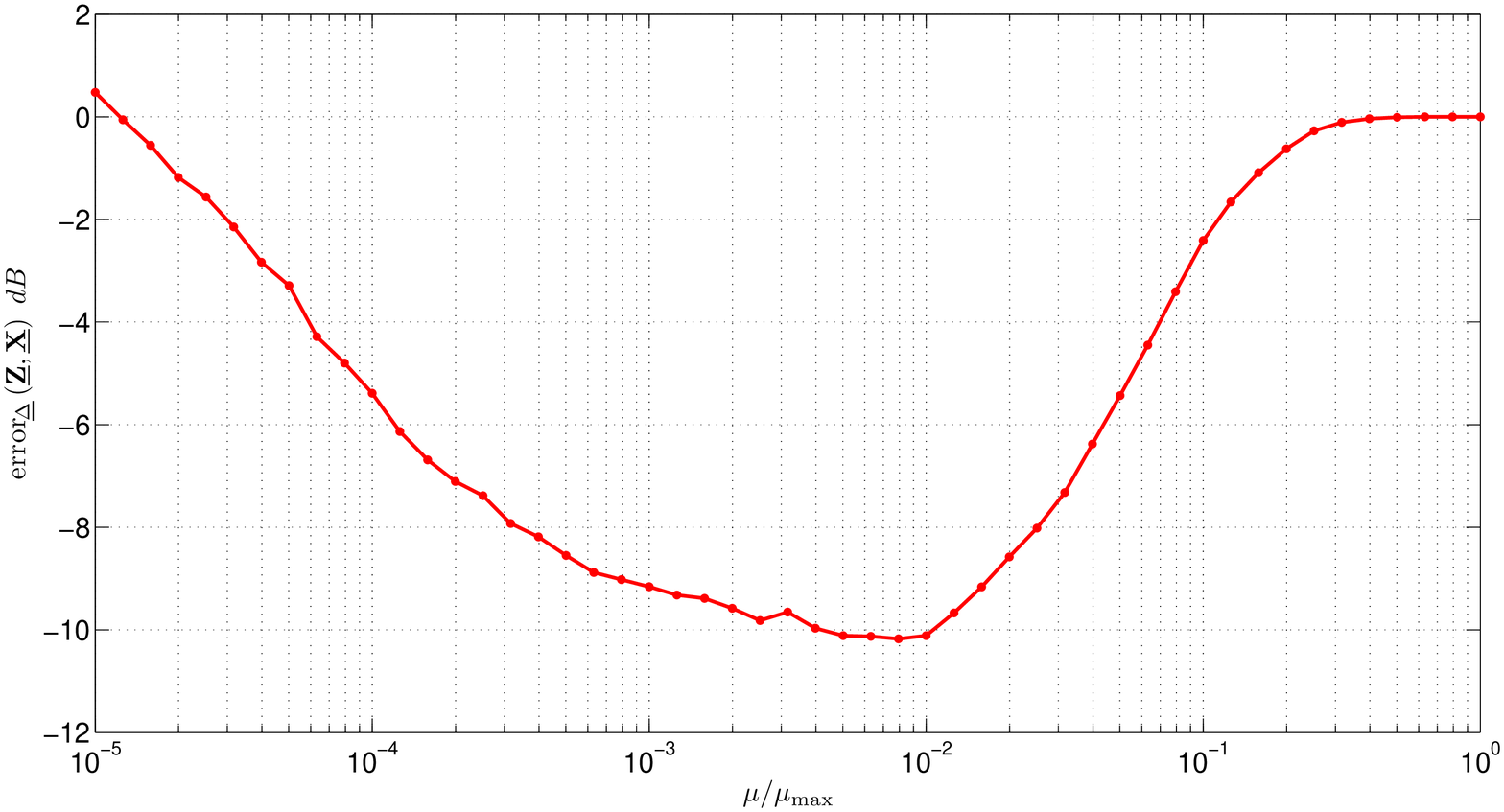}
  \caption{Performance of the low-rank tensor imputation method as function of the regularizing parameter $\mu$; (top) rank of the  tensor as recovered by \eqref{tensor_approximation} averaged over 100 test repetitions, (bottom) relative recovery error.}  \label{fig:rank_error_gauss}
\end{figure}

The relative recovery error between $\hat{\ZZ}$ and data $\ZZ$ was computed, along with the  rank of the recovered tensor, as a measure of performance.
Fig. \ref{fig:rank_error_gauss} depicts these figures of merit averaged over $100$ repetitions of the experiment, across  values of  $\mu$ varying on the interval $10^{-5}\mu_{\max}$ to $\mu_{\max}$, which is  computed as in Corollary \ref{corollary:mu_max}.
Fig \ref{fig:rank_error_gauss} (bottom) shows that the LRTI algorithm is successful in recovering the missing entries of $\ZZ$  up to $-10$dB for a wide range of values of $\mu$, presenting a minimum at $\mu=10^{-2}\mu_{\max}$. This result is consistent with   Fig. \ref{fig:rank_error_gauss} (top), which shows that  rank $R^*=6$ is approximately recovered at the minimum error.  Fig. \ref{fig:rank_error_gauss} (top) also corroborates the low-rank inducing effect of \eqref{tensor_approximation}, with the recovered rank varying from the  upper bound $\bar R=NP=16$ to $R=0$, as $\mu$ is increased, and  confirms that the recovered tensor is null at $\mu_{\max}$ as asserted by Corollary \ref{corollary:mu_max}.

\subsection{Simulated Poisson data}
\label{ssec:synthetic}
The synthetic example just described was repeated for the low-rank Poisson-tensor model described in Section \ref{sec:poisson}. Specifically, tensor data of dimensions $M\times N\times P=16\times 4\times 4$ were generated according to the low-rank Poisson-tensor model of Section \ref{sec:poisson}. Entries of $\ZZ$ consist of realizations of Poisson random variables generated according to \eqref{poisson_model}, with means specified by entries of $\XX$. Tensor $\XX$  is again constructed as in \eqref{parafac} from factors $\bA$, $\bB$ and $\bC$ having $R=6$ columns, containing the absolute value of realizations of independent Gaussian random variables scaled to yield $\E[x_{mnp}]=100$.
Half of the entries of $\ZZ$ were considered missing to be recovered from the  remaining half. Method \eqref{poisson_map} was employed for recovery, as implemented by the LRPTI Algorithm, with   regularization  $\frac{\mu}{2}(\|\bA\|_F^2+\|\bB\|_F^2+\|\bC\|_F^2)$.

\begin{figure}
  \includegraphics[width=0.8\linewidth,height=0.4\linewidth]{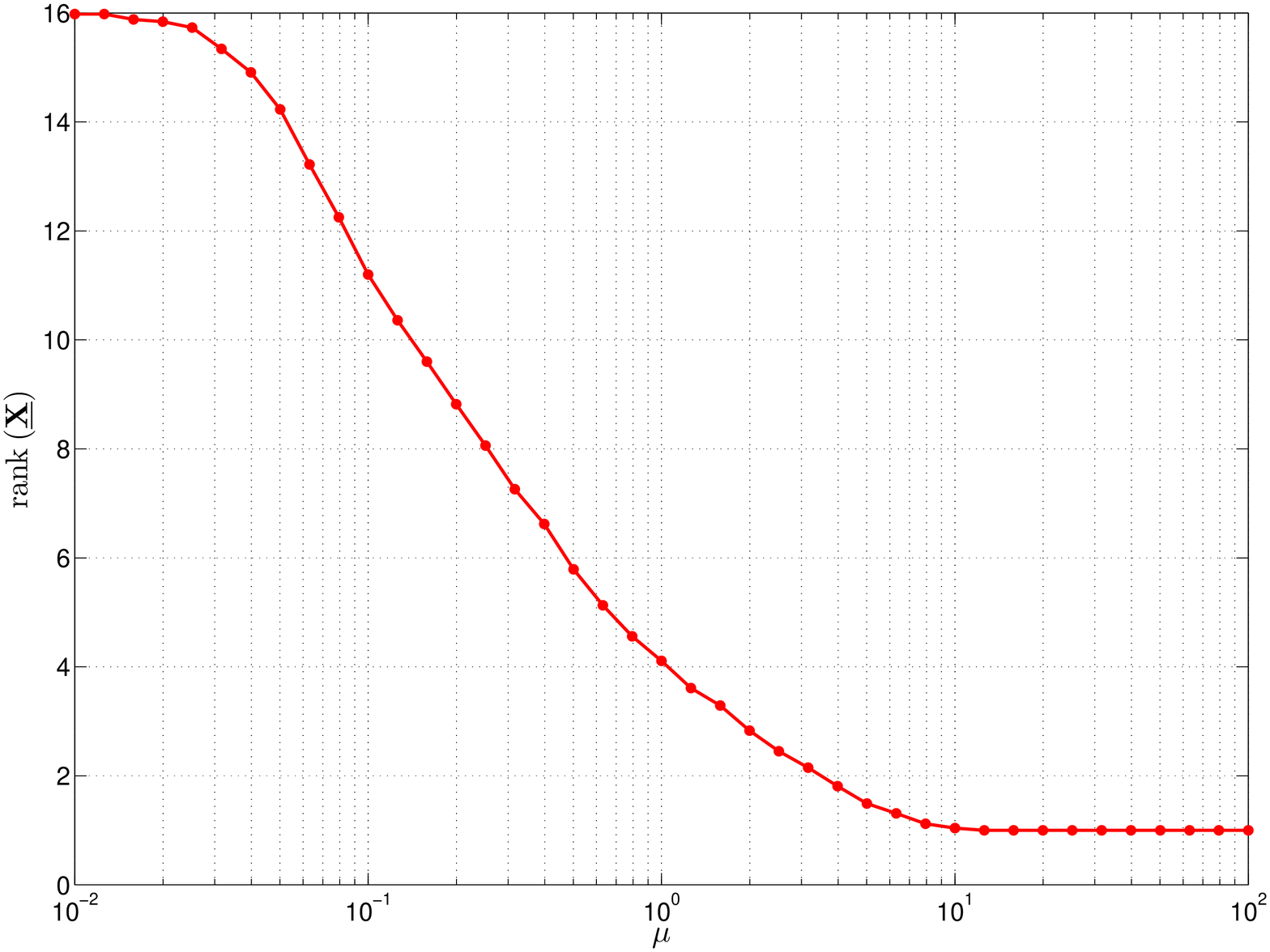}
   \includegraphics[width=0.8\linewidth,height=0.4\linewidth]{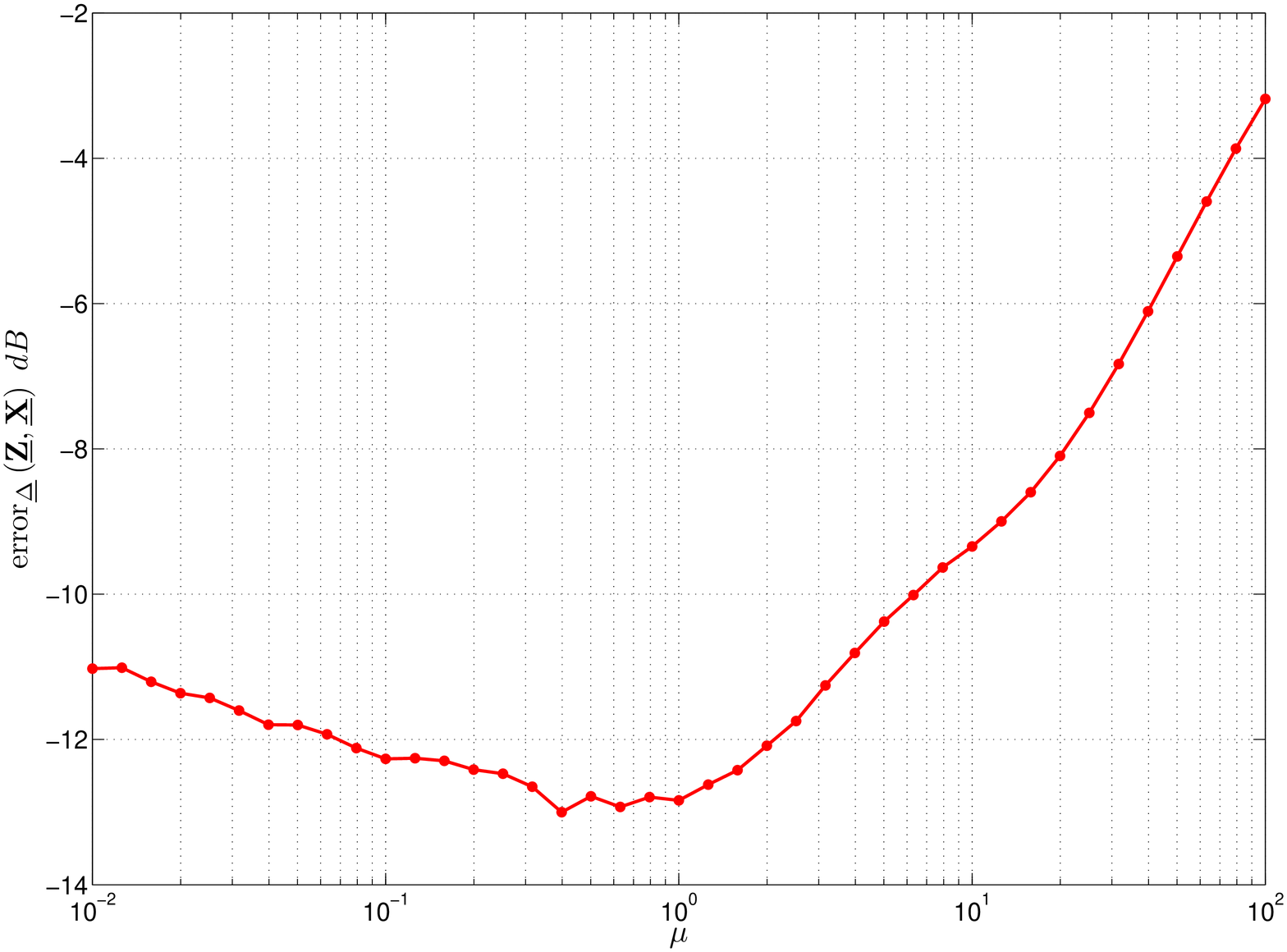}
  \caption{Performance of the low-rank Poisson imputation method as function of the regularizing parameter $\mu$; (top) rank of the recovered tensor averaged over 100 test repetitions, (bottom) relative recovery error.}  \label{fig:rank_error_poisson}
\end{figure}

Fig. \ref{fig:rank_error_poisson} shows the estimated rank and recovery error over $100$ realizations of the experiment, for $\mu$ in the interval $0.01$ to $100$.
The recovery error in Fig. \ref{fig:rank_error_poisson} (bottom) exhibits a minimum of $-15$dB at  $\mu=1$, where the   rank $R^*=6$ is recovered [cf. Fig. \ref{fig:rank_error_poisson} (top).]   The   low-rank inducing effect of \eqref{tensor_approximation} is again corroborated by the decreasing trend in Fig. \ref{fig:rank_error_poisson} (top), but in this case the rank is lower bounded by $R=1$, because the K-L fitting criterion prevents \eqref{poisson_map} from yielding a null estimate $\hat{\ZZ}$.

\subsection{MRI data}

\begin{figure}[t]
\begin{minipage}[b]{.48\linewidth}
  \centering
\centerline{\epsfig{figure=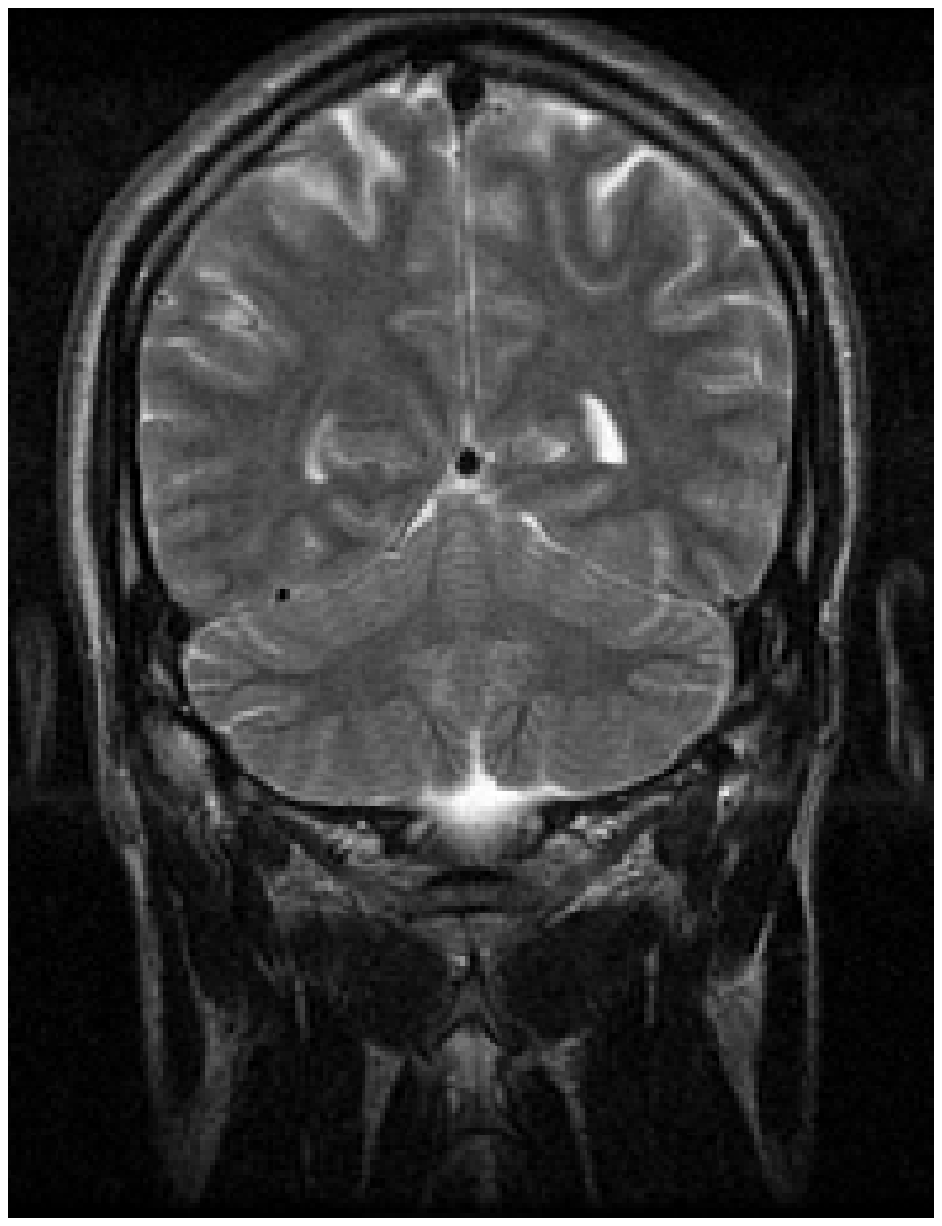,width=4.0cm,height=4cm}}
%  \vspace{1.5cm}
  %\centerline{(b) Results 3}\medskip
\end{minipage}
\hfill
\begin{minipage}[b]{0.48\linewidth}
  \centering
  \centerline{\epsfig{figure=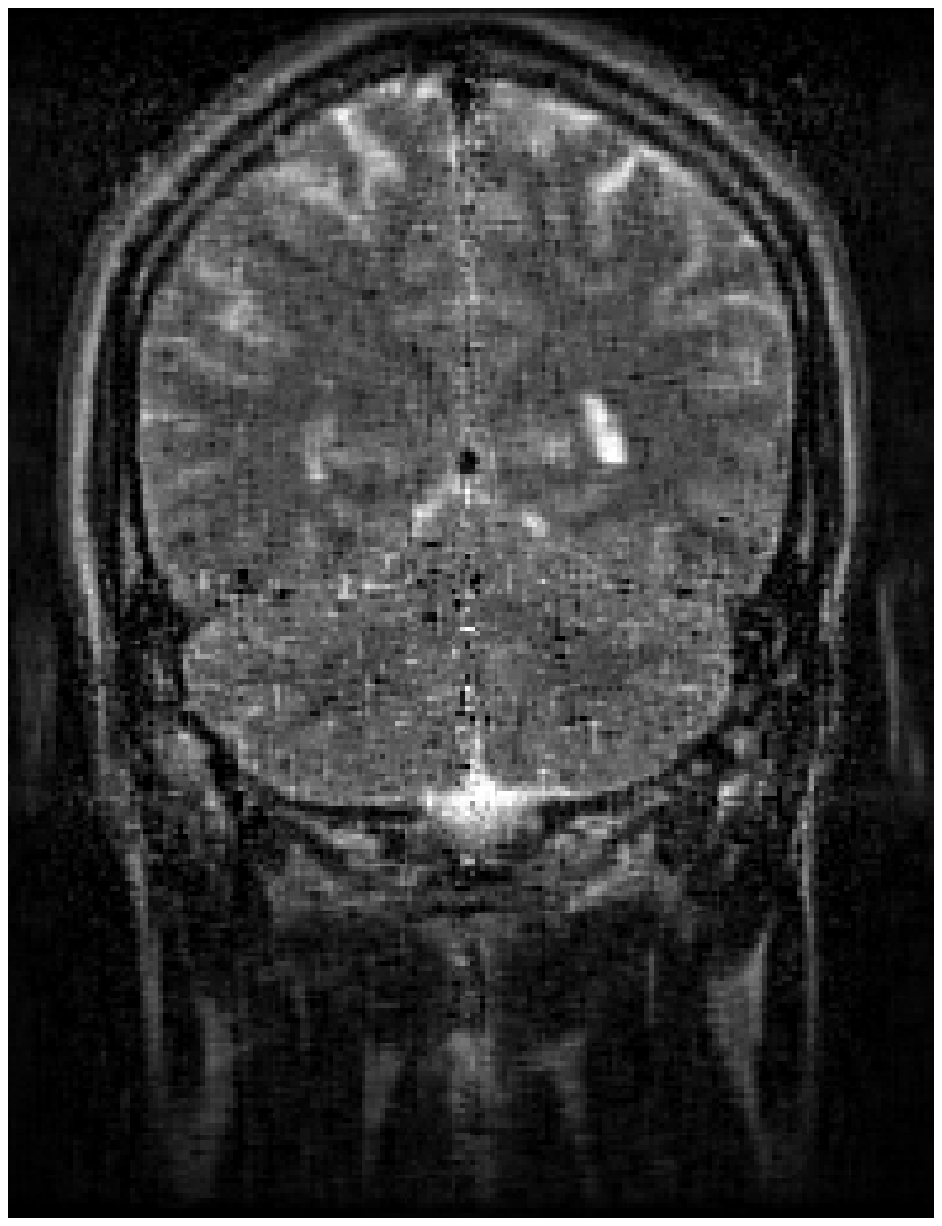,width=4.0cm,height=4cm}}
%  \vspace{1.5cm}
 % \centerline{(c) Result 4}\medskip
\end{minipage}

\vspace{.5cm}
\begin{minipage}[b]{.48\linewidth}
  \centering
  \centerline{\epsfig{figure=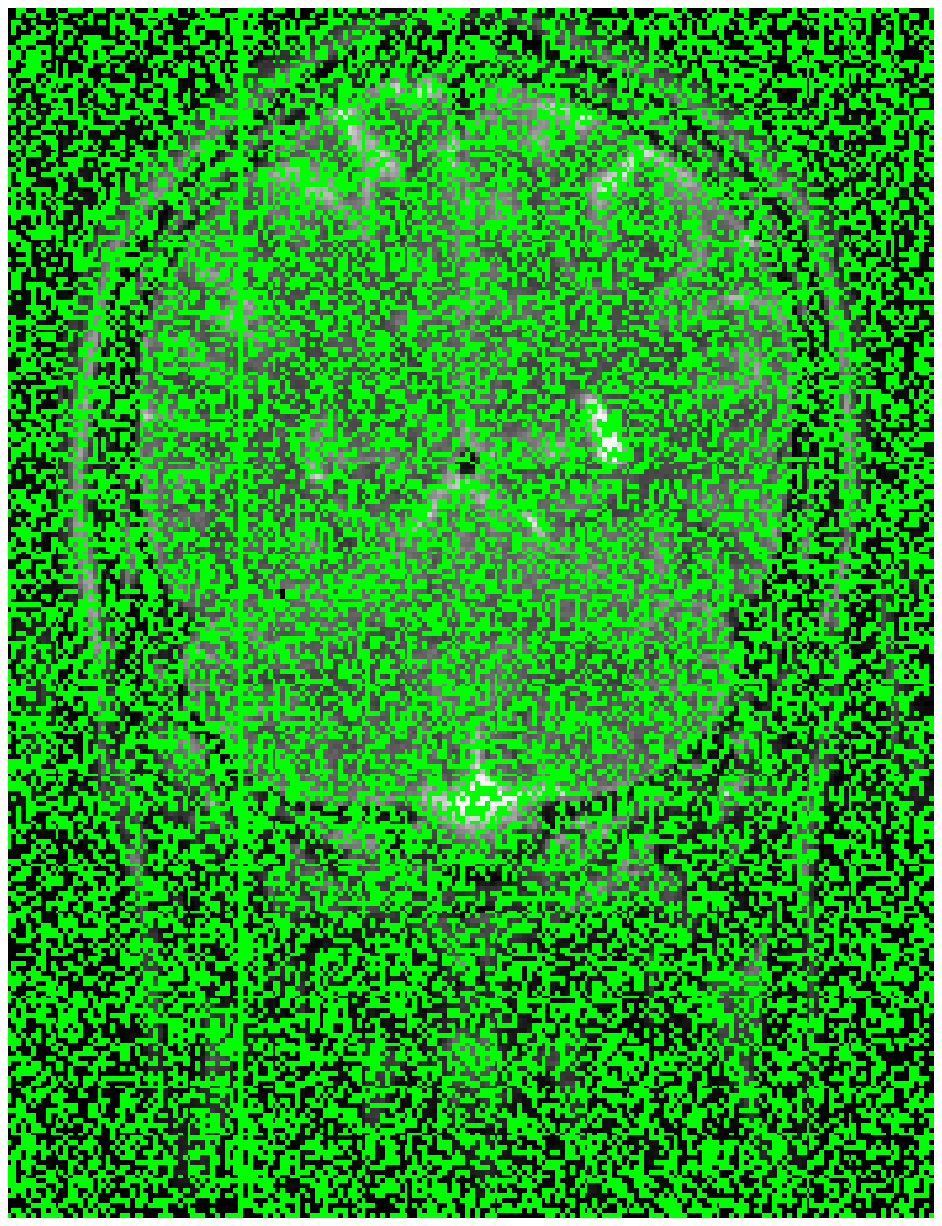,width=4cm,height=4cm}}
%  \vspace{1.5cm}
  %\centerline{(b) Results 3}\medskip
\end{minipage}
\hfill
\begin{minipage}[b]{0.48\linewidth}
  \centering
  \centerline{\epsfig{figure=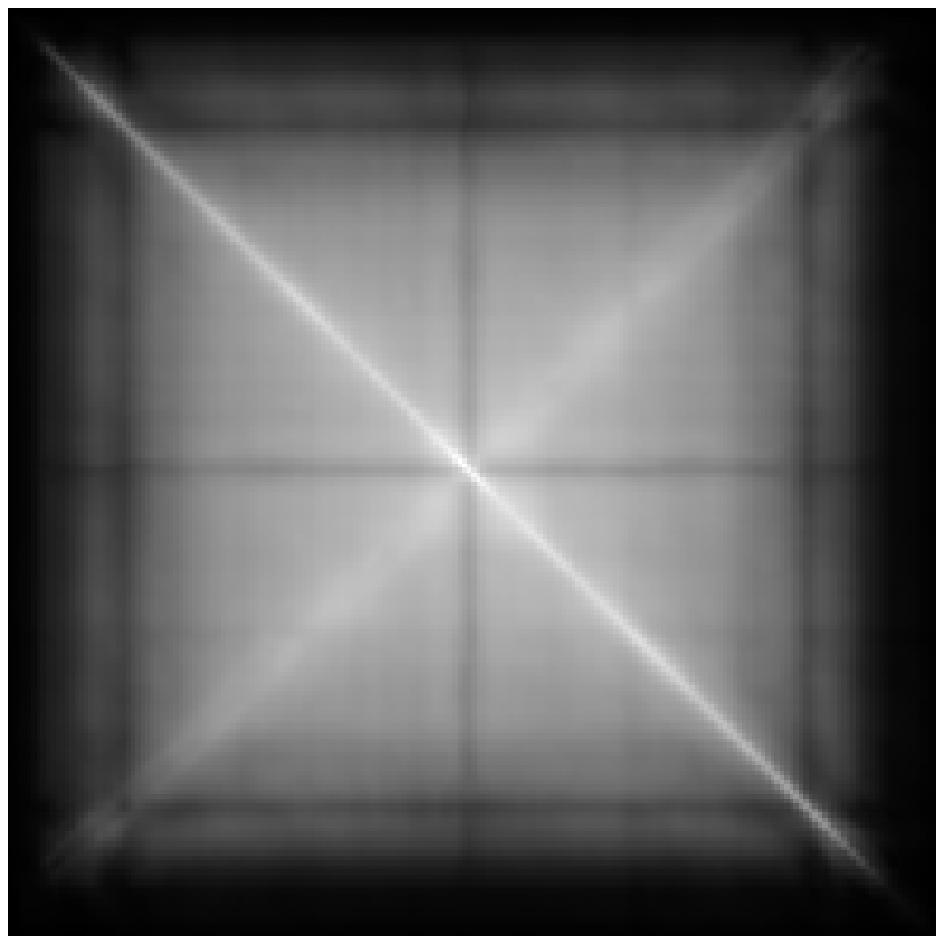,width=4cm,height=4cm}}
%  \vspace{1.5cm}
  %\centerline{(c) Result 4}\medskip
\end{minipage}

\vspace{.5cm}
\begin{minipage}[b]{.48\linewidth}
  \centering
  \centerline{\epsfig{figure=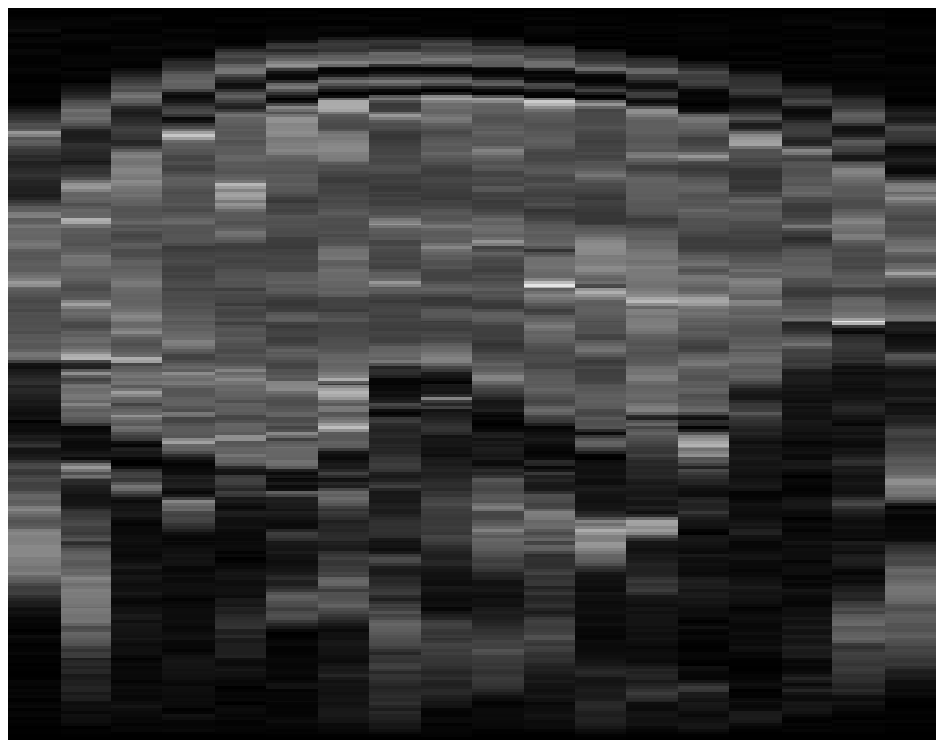,width=4cm,height=4cm}}
%  \vspace{1.5cm}
 % \centerline{(b) Results 3}\medskip
\end{minipage}
\hfill
\begin{minipage}[b]{0.48\linewidth}
  \centering
  \centerline{\epsfig{figure=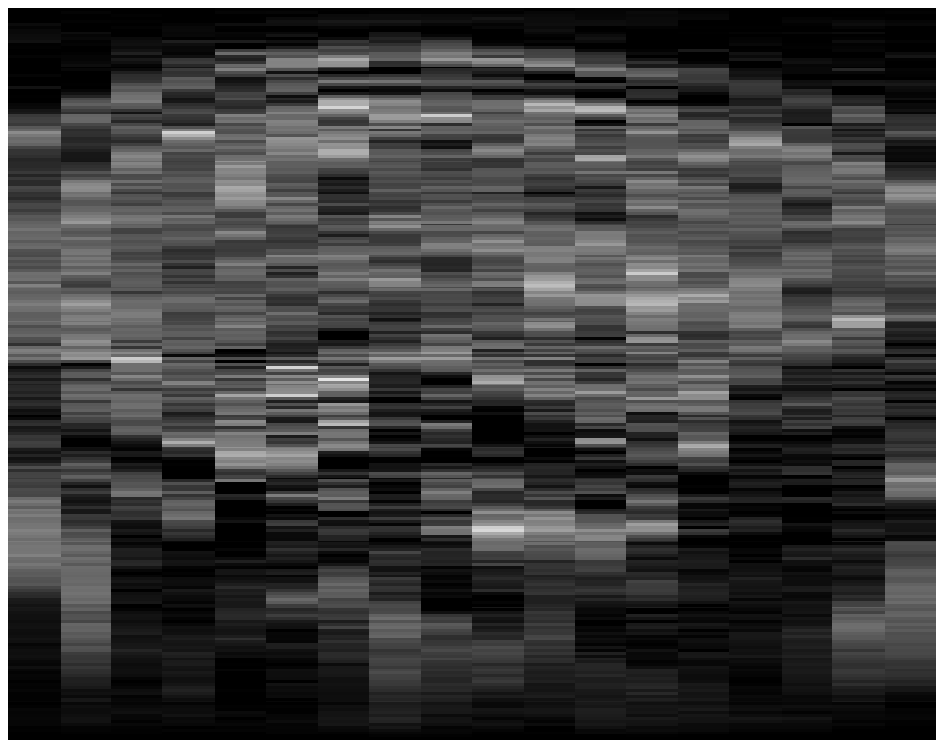,width=4cm,height=4cm}}
%  \vspace{1.5cm}
%  \centerline{(c) Result 4}\medskip
\end{minipage}
\caption{Results of applying \eqref{kernel_estimator} to the MRI brain data set 657. (top) original and recovered fibers $\bZ_p$ and $\hat{\mathbf{Z}}_p$ for $p=5$. (center) input fiber $\bZ_{p}, \ p=5$ with missing data, and covariance matrix $\bK_{\caliN}$. (bottom) original and recovered columns $\bZ_n$ and $\hat{\mathbf{Z}}_n$ for the position  $n=50$ in which the whole input   slice is missing ) }
\label{fig:res}
\end{figure}

Estimator \eqref{kernel_estimator} was tested against a corrupted version of the
MRI brain data set 657 from the Internet brain segmentation repository~\cite{ibsr}. The tensor $\ZZ$ to
be estimated corresponds to a three-dimensional MRI scan of the brain comprising
a set of $P=18$ images, each of $M\times N=256\times 196$ pixels. Fifty  percent of the
data is removed uniformly at random together with  the whole slice
$\bZ_n,\ n=50$. Fig. \ref{fig:res} depicts the
results of applying  estimator \eqref{kernel_estimator} to the remaining data, which yields a reconstruction error of $-10.54$dB. The original slice $\bZ_p,\ p=5$,  its corrupted counterpart, and the
resulting estimate are shown on top and center left.  Covariance matrices  $\bK_{\caliM}$,  $\bK_{\caliN}$  and
$\bK_{\caliP}$ are estimated  from six  additional tensor samples containing complementary scans of the brain also available at \cite{ibsr}. Fig. \ref{fig:res} (center right) represents the covariance matrix $\bK_{\caliN}$ for column slices perpendicular to $\bZ_p$, showing a structure that reflects symmetries of the brain.
This correlation is the key enabler for the method to recover the missing slice up to $-9.60$dB (see Fig. \ref{fig:res} (bottom)) by interpolating its a priori similar parallel counterparts.

All in all, the
experiment evidences the merits of low-rank PARAFAC decomposition for
modeling a tensor, the ability of the Bayesian estimator \eqref{tensor_map}
in recovering missing data, and the usefulness of
incorporating correlations as  side information.

On account of the comprehensive analysis of three-way MRI data arrays in \cite{czpa2009book}, and the nonnegative PARAFAC decomposition advanced thereby, inference of tensors with nonnegative continuous entries will be pursued as future research, combining methods and algorithms in sections \ref{sec:Bayesian} and \ref{sec:poisson} of this paper.

\subsection{RNA sequencing data}
\label{ssec:rna}
\begin{figure}[t]
\centering
  \begin{minipage}[b]{0.6\linewidth}
    \centering
    \centerline{\includegraphics[width=\linewidth]{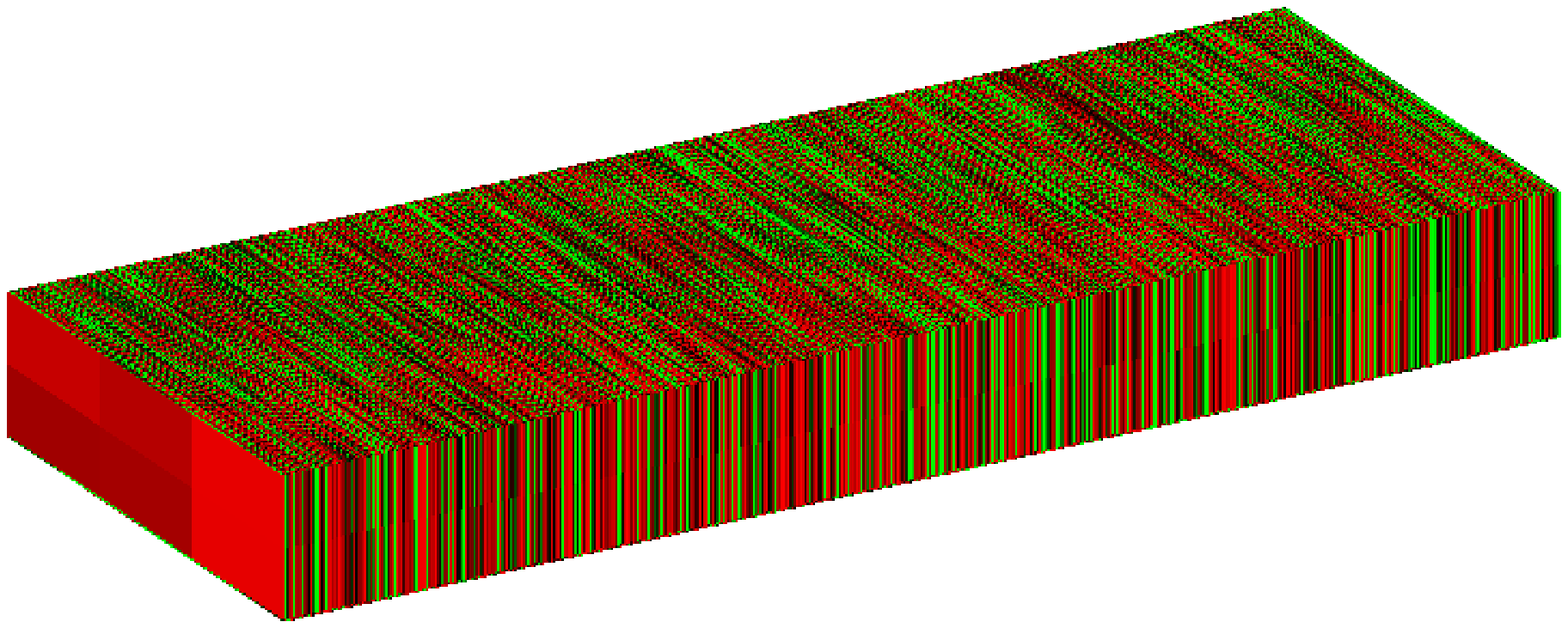}}
  \medskip
  \end{minipage}

  \begin{minipage}[b]{0.6\linewidth}
    \centering
    \centerline{\includegraphics[width=\linewidth]{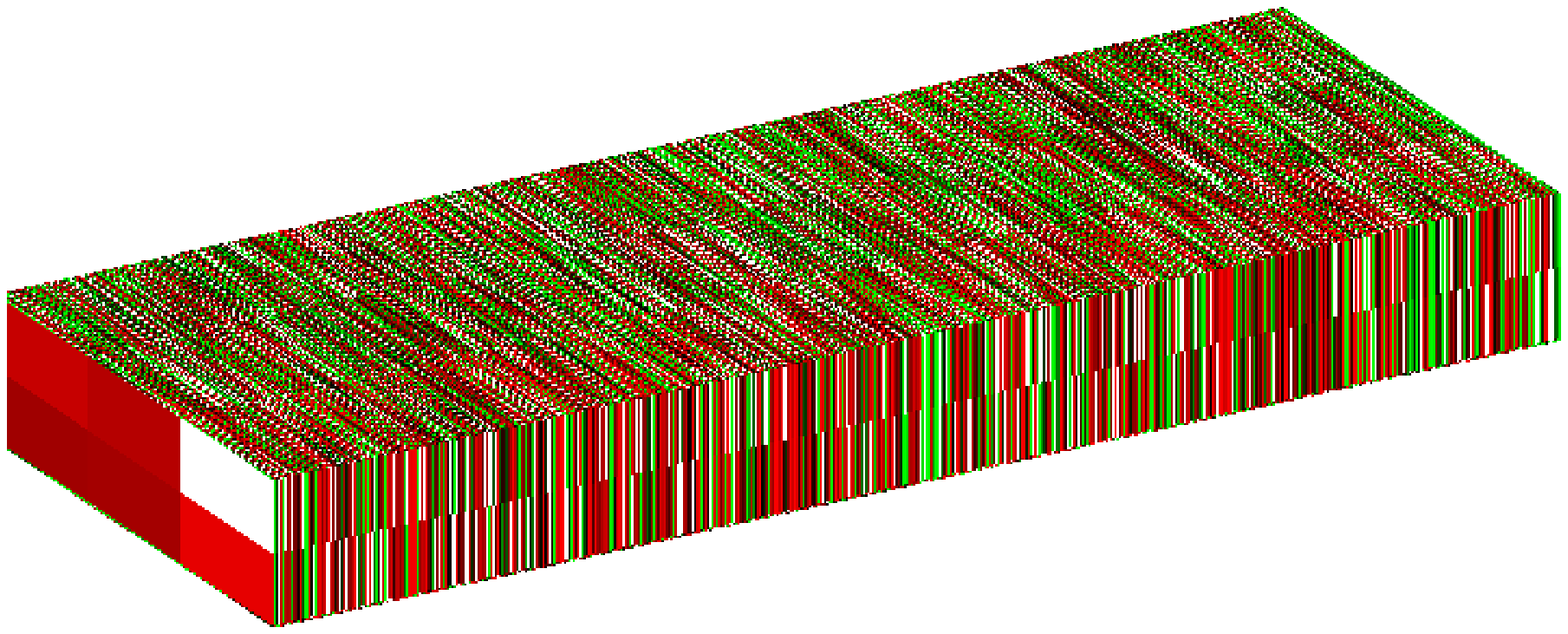}}
  \medskip
  \end{minipage}

    \begin{minipage}[b]{0.6\linewidth}
      \centering
      \centerline{\includegraphics[width=\linewidth]{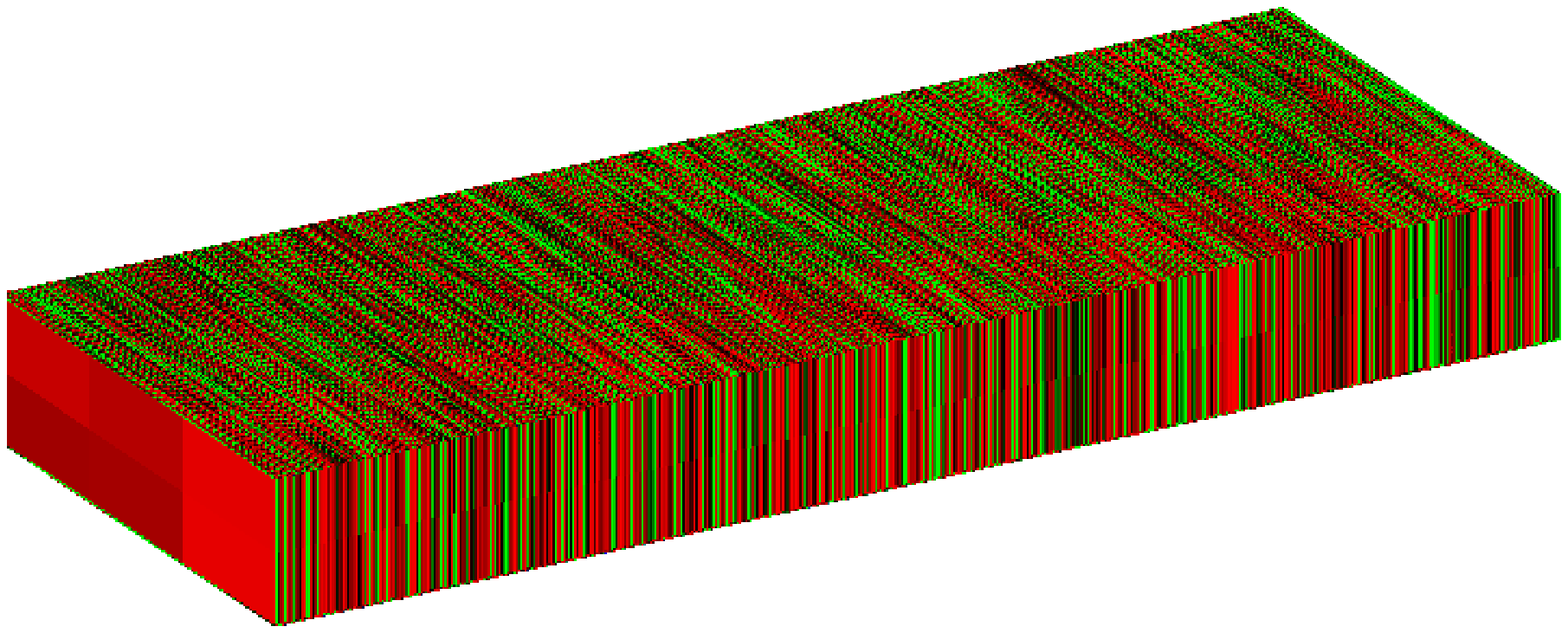}}
    \medskip
    \end{minipage}
%\center  \includegraphics[width=.5\linewidth]{figures/oblea.eps}\\

 %   \includegraphics[width=.5\linewidth]{figures/oblea_blanca15.eps}\\
%\vspace{-.1cm}
 %     \includegraphics[width=.5\linewidth]{figures/oblea_poisson15.eps}
  \caption{Imputation of sequencing count data via LRPTI; (top) original data; (center) data with missing entries; (bottom recovered tensor.}  \label{fig:obleas}
\end{figure}

The RNA-Seq method described in \cite{N08} exhaustively counts the number of
RNA transcripts from yeast cells.  The reverse transcription of RNA molecules into
cDNA is achieved by $P=2$ alternative methods, differentiated by the use of oligo-dT,
or random-hexonucleotide primers. These cDNA molecules are sequenced to obtain counts of
RNA molecules across $M=6,604$ genes  on the yeast genome. The experiment was repeated in \cite{N08} for a biological and a technological
replicate of the original sample totalling $N=3$ instances per primer selection. The data are
thus organized in a tensor of dimensions $6,604\times3\times2$ as shown in Fig. \ref{fig:obleas} (top),
with integer data that are modeled as  Poisson counts. Fifteen percent of the data is
removed and reserved for assessing performance. The missing data are  represented  in white in Fig. \ref{fig:obleas} (center).

The LRPTI algorithm is run with the data available in Fig. \ref{fig:obleas}  (center) producing the recovered tensor depicted in Fig. \ref{fig:obleas} (bottom). The recovery error for this experiment was $-15$dB.

\section{Concluding summary}\label{sec:conc}
It was shown in this paper that regularizing with the  Frobenius-norm square of the PARAFAC decomposition factors, controls the tensor's rank by inducing sparsity in the vector of amplitudes of its rank-one components.
A Bayesian method for tensor completion was developed based on this property, introducing priors on the tensor factors. It was argued, and corroborated numerically,  that this prior information endows the completion method with extra capabilities in terms of smoothing and  extrapolation. It was also suggested through a parallelism between Bayesian and RKHS inference, that the prior covariance matrices can be obtained from (sample) correlations among the tensor's slices.
In such a probabilistic  context, generic distribution models for the data lead to multiple fitting criteria. Gaussian and Poisson processes were especially considered by developing algorithms that minimize the regularized LS and K-L divergence, respectively.

Numerical tests on synthetic data corroborated the low-rank inducing property, and the ability of the completion method to recover the ``ground-truth'' rank, while experiments with brain images and gene expression levels in yeast served to evaluate the method's
 performance on real datasets.

Although the results and algorithms  in this paper were presented for three-way arrays, they are readily extendible to higher-order tensors or reducible to the matrix case.

\appendix
\textbf{I. Proof of Proposition \ref{prop:prop_matrix}}
\begin{IEEEproof}
\textbf{a)}
The equivalence of  \eqref{nuclearnorm} and \eqref{nuclear_norm_lr} results immediately from \eqref{nuclear_norm_decomposition}. Indeed,
if  \eqref{nuclear_norm_lr} is minimized in two steps
\begin{align}
\nonumber&\min_{\bX}\frac{1}{2} \min_{\mathbf{B},\mathbf{C}}
\|{\bf(Z-X)}\hadprod \bDelta\|_F^2+\frac{\mu}{2}(\|\mathbf
C\|_F^2+\|\mathbf B\|_F^2)\\
 &\textrm{ s. to }\mathbf C\mathbf{B}^T=\mathbf{X} \label{nuclearnorm_twosteps}
\end{align}
 it is apparent that the LS part of the cost does not depend on the inner minimization variables. Hence, \eqref{nuclearnorm_twosteps} can be rewritten as
\begin{align}
\min_{\bX}
\frac{1}{2}\|{\bf(Z-X)}\hadprod \bDelta\|_F^2+ \hspace{-1cm}\min_{\substack{\mathbf{B},\mathbf{C}\\ \textrm{\hspace{1cm} s. to }\mathbf C\mathbf{B}^T=\mathbf{X}}} \hspace{-.5cm} \frac{\mu}{2}(\|\mathbf
C\|_F^2+\|\mathbf B\|_F^2) \label{nuclearnorm_almost}
\end{align}
and by recognizing \eqref{nuclear_norm_decomposition} as the inner problem in \eqref{nuclearnorm_almost}, the equivalence follows.

\textbf{b)}
Consider the cost in \eqref{nuclear_norm_lr} at the local minimum $(\boB,\boC)$
\begin{align*}
U(\boB,\boC):=\frac{1}{2}\|(\bZ-\boX )\hadprod\bDelta\|_F^2+\frac{\mu}{2}(\|\boC\|_F^2+\|\boB\|_F^2) %\label{costU}.
\end{align*}
where $\boX:=\boB\boC^T$. Arguing  by contradiction, suppose that there is a different local minimum $(\bB,\bC)$ such that $U(\bB,\bC)\neq U(\boB,\boC)$, and without loss of generality set $U(\bB,\bC)< U(\boB,\boC)$, so that
$dU:= U(\bB,\bC)-U(\boB,\boC)<0$,
which can be expanded to
\begin{align}
\nonumber dU&= \trace\left[\left(\bDelta\hadprod( \bZ-\boX)\right)\left(\bDelta\hadprod( \boX-\bX)\right)\right]+\|\bDelta\hadprod( \boX-\bX)\|_F^2\\
&+\frac{\mu}{2}\left(\|\bC\|_F^2-\|\boC\|_F^2+\|\bB\|_F^2-\|\boB\|_F^2\right)<0.\label{deltaU}
\end{align}

Setting this inequality aside for now, consider the augmented matrix $\bQ$ in terms of generic $\bB$ and $\bC$ matrices:
\begin{align}
\bQ:=\left[
       \begin{array}{c}
         \bB \\
         \bC \\
       \end{array}
     \right]\left[
              \begin{array}{cc}
                \bB^T & \bC^T \\
              \end{array}
            \right]=\left(
                      \begin{array}{cc}
                        \bB\bB^T & \bX \\
                        \bX^T & \bC\bC^T \\
                      \end{array}
                    \right)\label{defQ}
\end{align}
and the corresponding $\boQ$  defined in terms of $\boB$ and $\boC$.

For each value of $\theta\in (0,1)$ consider the convex combination
\begin{align}
\bQ_\theta:=\boQ +\theta (\bQ - \boQ).\label{Qtheta}
\end{align}
As both  $\bQ$ and $\boQ$ are positive semi-definite, so is $\bQ_\theta$ and by means of the Choleski factorization one obtains
\begin{align}
\bQ_\theta:=\left[
       \begin{array}{c}
         \bB_\theta \\
         \bC_\theta \\
       \end{array}
     \right]\left[
              \begin{array}{cc}
                \bB_\theta' & \bC_\theta' \\
              \end{array}
            \right]=\left(
                      \begin{array}{cc}
                        \bB_\theta\bB_\theta' & \bX_\theta \\
                        \bX_\theta' & \bC_\theta\bC_\theta' \\
                      \end{array}
                    \right).\label{choleski}
\end{align}
which defines $\bB_\theta$, $\bC_\theta$ and $\bX_\theta$.

Expanding the cost difference $dU_\theta$ as in \eqref{deltaU}  results in
\begin{align*}
dU_\theta&:= U(\bB_\theta,\bC_\theta)-U(\boB,\boC)\\
 &=\trace\left[\left(\bDelta\hadprod( \bZ-\boX)\right)\left(\bDelta\hadprod( \boX-\bX_\theta)\right)\right]\\
&+\frac{\mu}{2}\left(\|\bC_\theta\|_F^2-\|\boC\|_F^2+\|\bB_\theta\|_F^2-\|\boB\|_F^2\right) \\
&+\|\bDelta\hadprod( \boX-\bX_\theta)\|_F^2.%\label{deltaUtheta}
\end{align*}

From the definitions \eqref{defQ}-\eqref{choleski}  it follows that $\boX-\bX_\theta=\theta(\boX-\bX)$, $\|\bB_\theta\|_F^2-\|\boB\|_F^2=\theta(\|\bB\|_F^2-\|\boB\|_F^2)$, and   $\|\bC_\theta\|_F^2-\|\boC\|_F^2=\theta(\|\bC\|_F^2-\|\boC\|_F^2)$, so that
\begin{align*}
dU_\theta&:= \theta \trace\left[\left(\bDelta\hadprod( \bZ-\boX)\right)\left(\bDelta\hadprod( \boX-\bX)\right)\right]\\
&+\frac{\mu\theta}{2}\left(\|\bC\|_F^2-\|\boC\|_F^2+\|\bB\|_F^2-\|\boB\|_F^2\right)\\
&+\theta^2\|\bDelta\hadprod( \boX-\bX_\theta)\|_F^2% \label{deltaUthetatheta}
\end{align*}
and thus, it can be put in terms of \eqref{deltaU} as in
\begin{align*}
dU_\theta&:= \theta \left(dU-\|\bDelta\hadprod( \boX-\bX_\theta)\|_F^2\right) +\theta^2\|\bDelta\hadprod( \boX-\bX_\theta)\|_F^2 .%\label{deltatheta}
\end{align*}

If $dU$ were strictly negative, so would  $dU-\|\bDelta\hadprod( \boX-\bX_\theta)\|_F^2$,  and  hence
\begin{align*}\lim_{\theta\rightarrow 0} \frac{1}{\theta}dU_\theta=\left(dU-\|\bDelta\hadprod( \boX-\bX_\theta)\|_F^2\right)<0.
\end{align*}
but then there is in every neighborhood of $(\boB,\boC)$ a point $(\bB_\theta,\bC_\theta)$ such that $U(\bB_\theta,\bC_\theta)<U(\boB,\boC)$, $\boB,\boC$ cannot be a local minimum. This contradiction  implies  that $U(\bB,\bC)=U(\boB,\boC)$  for any pair of local minima, which proves the statement in part b) of Proposition \ref{prop:prop_matrix}.
\end{IEEEproof}
\textbf{II-Equivalence of tensor completion problems}
\begin{IEEEproof}
The Frobenius square-norms of $\bA$, $\bB$, and $\bC$  are separable across columns; hence, the penalty in \eqref{tensor_approximation} can be rewritten as
\begin{align}
\nonumber\|\bA\|_f^2+\|\bB\|_F^2+\|\bC\|_F^2&=\sum_{r=1}^R \|\mathbf a_r\|^2+\|\mathbf  b_r\|^2+\|\mathbf  c_r\|^2\\
&=\sum_{r=1}^R  a_r^2+ b_r^2+ c_r^2\label{separable_penalty}
\end{align}
by defining $a_r:=\|\mathbf a_r\|$, $b_r:=\|\mathbf  b_r\|$ , $c_r:=\|\mathbf  c_r\|$, $r=1,\ldots,R$.

On the other hand, $\XX$ can be expressed w.l.o.g. in terms of the normalized outer products
\eqref{normalized_outerproduct} with $\gamma_r:=a_r  b_r c_r$. Substituting \eqref{normalized_outerproduct} and \eqref{separable_penalty} for the tensor and the penalty respectively, \eqref{tensor_approximation} reduces to
\begin{align}\label{atomic norm_proof}
\nonumber \min_{\{\mathbf{\hat u}\},\{\mathbf{\hat v\}},\{\mathbf{\hat w}\}}\min_{\bm\gamma}\min_{\{a_r\},\{b_r\},\{c_r\}}&
\frac{1}{2}||\left(\ZZ-\XX\right)\hadprod  \DD||_F^2\\
\nonumber&+\frac{\mu}{2}\sumr a_r^2+ b_r^2+ c_r^2 \\
&\hspace{-0cm}{\rm s. \: to\;}\mathbf
\XX=\sum_{r=1}^R  \gamma_r (\mathbf u_r\circ\mathbf v_r\circ\mathbf w_r)\nonumber\\
&\hspace{-0cm}\gamma_r= a_r  b_r c_r.
\end{align}

Focusing  on the  inner minimization  w.r.t. norms $a_r$, $ b_r$, and $  c_r$  for arbitrary fixed directions $\{\mathbf{u}_r\}$, $\{\mathbf{v}_r\}$, and $\{\mathbf{w}_r\}$, and fixed products  $\gamma_r:= a_r  b_r  c_r$. The constraints and hence the LS part of the cost depend on $\gamma_r$ only, and not on their particular factorizations  $a_r  b_r  c_r$. Thus, the penalty is the only term that varies when $\gamma_r$ is constant, rendering the inner-most minimization in \eqref{atomic norm_proof}  equivalent to
\begin{align}
\min_{ a_r,  b_r,  c_r}&
 \nonumber  a_r^2+  b_r^2+  c_r^2 \\
\gamma_r= a_r  b_r  c_r.
\label{atom norm_abc}\end{align}

The arithmetic geometric-mean inequality gives the solution to \eqref{atom norm_abc}, as it states that for  scalars $a_r^2$, $b_r^2$ and $b_r^2$, it holds that
\begin{align*}%\label{geometric_artitmetic_mean}
\sqrt[3]{a_r^2b_r^2c_r^2}\leq (1/3)(a_r^2+b_r^2+c_r^2)
\end{align*}
with equality when $a_r^2=b_r^2=c_r^2$, so that the minimum of \eqref{atom norm_abc} is attained at $a_r^2=b_r^2=c_r^2=\gamma_r^{2/3}$.

Substituting the corresponding $\sumr (a_r^2+b_r^2+c_r^2)=3\sumr \gamma_r^{2/3} =3\|\bm\gamma\|_{2/3}^{2/3}$ into \eqref{atomic norm_proof} yields \eqref{atomic norm_23}. Equivalence of the optimization problems is transitive; hence, by showing that both  \eqref{atomic norm_23} and  \eqref{tensor_approximation}  equivalent to   \eqref{atomic norm_proof} proves them equivalent to each other, as desired.
\end{IEEEproof}

\textbf{III. Proof of Corollary \ref{corollary:mu_max}}
\begin{IEEEproof}
The following result  on the norm of the matrix inverse will be used in the proof of the corollary.

\begin{lemma}\cite[p.58]{golub}
If $\mathbf E\in R^{m\times m}$  satisfies $\|\mathbf E\|_F\leq 1$, then $\mathbf I+\mathbf E$ is invertible, and  $\left\|(\mathbf I+\mathbf E)^{-1}\right\|_F\leq (1-  \|\mathbf E\|_F)^{-1}.$
\end{lemma}

Another useful inequality  holds for any value of $\mu$, and for $\bA$, $\bB$, and $\bC$ being the minimizers of \eqref{tensor_approximation}
\begin{align}
\mu \left(\|\bA\|_F^2+\|\bB\|_F^2+\|\bC\|_F^2\right)\leq \|\DD\hadprod \ZZ\|_F^2\label{bound_comparing20}
\end{align}
 as it follows from comparing the cost at such a minimum, and at the feasible point $(\bA,\bB,\bC)=(\mathbf 0,\mathbf 0,\mathbf 0)$.

A second characterization of the minimum of \eqref{tensor_approximation} will be obtained by equating the gradient to zero.
By vectorizing matrix $\bA$,  the cost in \eqref{tensor_approximation}  can be rewritten as
\begin{align}
\sum_{p=1}^P\frac{1}{2}\left\|\Diag[\bm\delta_p]\left(\mathbf z_p-(\bB\Diag[\mathbf e_p^T\bC]\otimes \mathbf I))\mathbf a\right)\right\|_2^2 +\frac{\mu}{2}\|\mathbf a\|^2_2\label{vectorized_nucleaqrnorm}
\end{align}
where $\mathbf z_p$, $\bm\delta_p$, and $\mathbf a$ denote the vector rearrangements  of matrices $\bZ_p$, $\bD_p$ , and $\bA$, respectively.
Additional  regularization that vanishes when taking derivatives w.r.t. $\bA$  were removed from \eqref{vectorized_nucleaqrnorm}.
Setting the gradient of \eqref{vectorized_nucleaqrnorm} w.r.t. $\mathbf a$ to zero, yields
\begin{align*}
\mathbf a&=(\mathbf I+\mathbf E)^{-1}\bm\zeta
\end{align*}
with
\begin{align*}
\mathbf E&:=\frac{1}{\mu}\sum_{p=1}^P \left(\bB^T\Diag[\mathbf e_p^T\bC]\otimes\mathbf I\right)\Diag[\bm\delta_p]\left(\bB\Diag[\mathbf e_p^T\bC]\otimes\mathbf I\right)\\
\bm\zeta &:=\frac{1}{\mu}\sum_{p=1}^P \left(\bB^T\Diag[\mathbf e_p^T\bC]\otimes\mathbf I\right)\Diag[\bm\delta_p] \mathbf z_p.
%\label{solve_for_a}.
\end{align*}
The norms of $\mathbf E$ and $\bm \zeta$ can be bounded by  using the sub-multiplicative property of the norm,  and the Cauchy-Schwarz inequality, which results in
\begin{align*}
 \|\mathbf E\|_F&\leq \frac{1}{\mu}\|\bB\|_F^2\|\bC\|_F^2\\
  \|\bm\zeta\|_F&\leq \frac{1}{\mu}\|\DD\hadprod\ZZ\|_F\|\bB\|_F\|\bC\|_F.
\end{align*}
Then according to the previous lemma, if $\mu$ is chosen large enough so that $\|\mathbf E\|_F\leq 1$ then the norm of  $\bA$ is bounded by
\begin{align}
\|\bA\|_F=\|\mathbf a\|_2\leq (\mu-\|\bB\|_F^2\|\bC\|_F^2)^{-1}\|\bB\|_F\|\bC\|_F\|\DD\hadprod\ZZ\|_F\label{inequality_stationary}
\end{align}
which constitutes the sought second characterization of the minimum of \eqref{tensor_approximation}.

Yet a third characterization  was obtained during the proof of Proposition \ref{proposition_2}, in which the norm of the factor columns  were shown equal to each other, so that
\begin{align}
\|\bA\|_F=\|\bB\|_F=\|\bC\|_F\label{equal_frob_norm}.
\end{align}

Substituting \eqref{equal_frob_norm} into \eqref{bound_comparing20} and  \eqref{inequality_stationary} yields
\begin{align}
 \|\bA\|_F^2 &\leq \|\DD\hadprod \ZZ\|_F^2/3\mu\label{bound_comparing20A}\\
\|\bA\|_F&\leq (\mu-\|\bA\|_F^4)^{-1}\|\bA\|^2_F\|\DD\hadprod\ZZ\|_F\label{inequality_stationaryA}.
\end{align}

Form \eqref{inequality_stationaryA}, two cases are found possible:
\begin{align}
\nonumber            \textbf{case 1: }&                                                                                                   \|\bA\|_F=\mathbf 0 \textrm{; and} \\
             \label{corollary_cases}             \textbf{case 2: }&                                                     1\leq (1-\|\bA\|_F^4/\mu)^{-1}\|\bA\|_F\|\DD\hadprod\ZZ\|_F/\mu.
                                                                          \end{align}

To argue that the second case is impossible, substitute   \eqref{bound_comparing20A} into \eqref{corollary_cases} and square the result to obtain
\begin{align}
1\leq (1-\|\DD\hadprod\ZZ\|_F^4/9\mu^3)^{-2}\|\DD\hadprod\ZZ\|_F^4/3\mu^3 \label{corollary_second_case}
\end{align}
But by hypothesis $\mu\geq \|\DD\hadprod\ZZ\|_F^{4/3}$ so that $\|\DD\hadprod\ZZ\|_F^4/\mu^3\leq 1$, and the right-hand side of \eqref{corollary_second_case} is bounded by $0.43$, so that the inequality does not hold. This implies that the first case in \eqref{corollary_cases}; i.e., $\|\bA\|_F=\mathbf0$, must hold, which in accordance with \eqref{equal_frob_norm}, further implies a null solution of \eqref{tensor_approximation}. That was the object of this proof. Still, the bound at $0.43$ can be pushed to one by  further reducing $\mu$, and the proof remains valid under the slightly relaxed condition  $\mu> (18/(5+\sqrt{21}))^{-1/3}\|\DD\hadprod\ZZ\|_F^{4/3}\simeq 0.81 \|\DD\hadprod\ZZ\|_F^{4/3}$.
\end{IEEEproof}
\textbf{IV-RKHS imputation}

Recursive application of Representer's Theorem yields
finite dimensional representations for the minimizers
$a_r$, $b_r$, and $c_r$ of \eqref{kernel_estimator}, given by
%
%$\hat a_{r}(x) ={\summ} \alpha_{r m} k_{\caliM}(x_m,x)$,  $\hat b_{r}(y) ={\sumn} \beta_{r n} k_{\caliN}(y_n,y),$ and $\hat c_{r}(t) %=\textstyle{\sump} \gamma_{r p} k_{\caliP}(t_p,t)$.
%
\begin{align*}
\hat a_{r}(m) &=\textstyle{\sum_{m'=1}^M} \alpha_{r m'} k_{\caliM}(m',m)\\
\hat b_{r}(n) &=\textstyle{\sum_{n'=1}^N} \beta_{r n'} k_{\caliN}(n',n)\\
\hat c_{r}(p) &=\textstyle{\sum_{p'=1}^P} \gamma_{r p'} k_{\caliP}(p',p).
\end{align*}
%
%\begin{align*}
%\hat a_{r}(x) &=\textstyle{\summ} \alpha_{r m} k_{\caliM}(x_m,x)
%\end{align*}
%%
%\begin{align*}
%\hat b_{r}(y) &=\textstyle{\sumn} \beta_{r n} k_{\caliN}(y_n,y)
%\end{align*}
%
%\begin{align*}
%\hat c_{r}(t) &=\textstyle{\sump} \gamma_{r p} k_{\caliP}(t_p,t)
%\end{align*}
%%
Defining vectors
$\bk_{\caliM}^T(m):=[k_{\caliM}(1,m),\ldots,k_{\caliM}(M,m)]$, and correspondingly
$\bk_{\caliN}^T(n):=[k_{\caliN}(1,n),\ldots,k_{\caliN}(N,n)]$, and $\bk_{\caliP}^T(p):=[k_{\caliP}(1,p),\ldots,k_{\caliP}(P,p)]$,
along with matrices  $\hat{\bA}\in \mathbb R^{M\times R}:\ \hat A(m,r):=\alpha_{m r}$,
$\hat{\bB}\in \mathbb R^{N\times R}:\ \hat B(n,r):=\beta_{n r}$, and $\hat{\bC}\in \mathbb R^{P\times R}:\ \hat C(p,r):=\gamma_{p r}$, it follows that
\begin{align}
\hat f_R(m,n,p){}={}&\sum_{r=1}^R \hat{a}_r(m) \hat{b}_{r}(n) \hat c_{r}(p)\nonumber\\
={}&\bk_{\caliM}^T(m) \hat{\bA} \Diag\left[\bk_{\caliP}^T(p)\hat{\bC}\right]\hat{\bB}^T
\bk_{\caliN}(n)\label{fR_kernel_appendix}.
\end{align}

Matrices $\hat{\bA},\ \hat{\bB}$, and $\hat{\bC}$ are further obtained
by solving
\begin{align*}
&\nonumber\min_{\hat{\bA},\hat{\bC},\hat{\bB}}
%\XX\in \mathbb^{M\times N\times P}   \hat{\bA}\in\mathbb
%R^{M\times R}\\  \hat{\bB}\in\mathbb R^{N\times R} \hat{\bC}\in\mathbb R^{P\times R}}  }
%\|(\ZZ-
%\XX)\hadprod \DD\|_F^2\\
\sump \left\|\left(\bZ_p-\bK_{\caliM} \bA \Diag\left[\bbept \bK_{\caliP}\bC\right]\bB^T\bK_{\caliN}\right)\hadprod \bDelta_p\right\|_F^2\\
\nonumber&\hspace{-0.04cm}+\hspace{-0.04cm}\frac{\mu}{2}\hspace{-0.04cm}
\left(\rm{trace}(\hspace{-0.04cm}\bA^T\hspace{-0.04cm}\bK_{\caliM}
\hspace{-0.04cm}\bA\hspace{-0.04cm})\hspace{-0.04cm}+\hspace{-0.04cm}\rm{trace}(\hspace{-0.04cm}\bB^T\hspace{-0.04cm}\bK_{\caliN} \hspace{-0.04cm}\bB\hspace{-0.04cm})\hspace{-0.04cm}+\hspace{-0.04cm}\rm{trace}(\hspace{-0.04cm}\bC^T\hspace{-0.04cm}\bK_{\caliP}\hspace{-0.04cm} \bC\hspace{-0.04cm})\right)\\
%&\textrm{s.\ to}\ \bX_p=\bK_{\caliM} \bA \bD_{(\bK_{\caliP}\bC),p}\bB^T\bK_{\caliN}\\
&\hspace{1cm}\hspace{-0.04cm}\textrm{s.\ to}\  \bA\in\mathbb R^{M\times R},\
\bB\in\mathbb R^{N\times R},\ \bC\in\mathbb R^{P\times R}%\label{ABC}
\end{align*}
which is transformed into \eqref{Kernel_tensor approximation} by changing variables $\bA=\mathbf \bK_{\caliM} \hat{\bA}$, $\bB=\mathbf \bK_{\caliN} \hat{\bB}$, and $\bC=\mathbf \bK_{\caliP} \hat{\bC}$, just as  \eqref{fR_kernel_appendix} becomes \eqref{fR_kernel}.

\textbf{V-Covariance estimation}

Inspection of  the entries of $\bK_{\caliP}(p,p'):=\E\left[\trace\left(\bX_p^T\bX_{p'}\right)\right]$ under the PARAFAC model, yields
\begin{align*}
\bK_{\caliP}(p,p')&:=\E\left[\trace\left(\sum_{r=1}^R\bbb_r \bbc_r(p)\bba_r^T\sum_{r'=1}^R\bba_{r'}\bbc_{r'}(p')\bbb_{r'}\right)\right]\\
&=\sum_{r=1}^R \sum_{r'=1}^R \E\left(\bbc^T_r(p) \bbc_{r'}(p') \right)\E\left( \bbb_{r'}^T\bbb_r\right)\E\left( \bba_r^T\bba_{r'}\right)\\
&=\sum_{r=1}^R \E\left(\bbc_r(p)\bbc_{r}(p')\right) \E \|\bbb_{r}\|^2\E \|\bba_{r}\|^2\\
&=\sum_{r=1}^R \bR_C(p,p') \trace(\bR_B) \trace(\bR_A)\\
&= R\ \bR_C(p,p') \trace(\bR_B) \trace(\bR_A)%\label{KXXP}
\end{align*}
which, after summing over $p'=p$, yields
\begin{align}
\nonumber \E\|\XX\|_F^2&=\sum_{p=1}^P \E\|\bX_p\|_F^2
=\sum_{p=1}^P \bR_{\caliP}(p,p)\\
&=R \trace(\bR_C) \trace(\bR_B) \trace(\bR_A)\label{Rttt}.
\end{align}

In addition, by incorporating the equal power assumption \eqref{equal_power}, equation \eqref{Rttt} further simplifies to
\begin{align*}
\E\|\XX\|_F^2=R\theta^3
\end{align*}
as stated in \eqref{ERtheta3}.
%CHAUCHAS

\textbf{VI - Vector form of \eqref{f_ls}}

 The $\textrm{vec}$ operator can be combined with the Kronecker product to factorize  $\textrm{vec}(\bA\bQ^T)=(\bQ \otimes \mathbf I)\textrm{vec}(\bA)$, and with the Hadamard product to convert it to a standard matrix product $\textrm{vec}(\bDelta\hadprod\bA)=\Diag(\textrm{vec}(\bDelta))\textrm{vec}(\bA)$. Using these two properties, \eqref{f_ls} can be put in terms of $\mathbf a:=\textrm{vec}(\bA)$ as in
\begin{align}
\nonumber f(\mathbf a)&:=\frac{1}{2}\sump ||\Diag(\textrm{vec}(\bDelta_p)) \left(\textrm{vec}(\bZ_p)-\bB\Diag({\mathbf e_P^T\bC})\mathbf a\right) ||_2^2\\
&+\frac{\mu}{2} \mathbf a (\mathbf I\otimes \bR_A^{-1}). \mathbf a\label{flsa_appendix}
\end{align}

\textbf{VII - Proof of Lemma \ref{lemma:majorizing}}
\begin{IEEEproof}
Function $g(\bA,\bar\bA)$ in \eqref{eq:major_g} is formed from $f(\bA)$ after substituting $g_1(\bA,\bar \bA)$ for $f_1(\bA)$, and $g_2(\bA,\bar \bA)$ for $f_2(\bA)$, respectively, as defined by
\begin{align}
f_1(\bA)&:=\trace\left(\mathbf A^T \bR_A ^{-1}\mathbf A\right)\label{f1prrof_lemma}\\
g_1(\bA,\bar\bA)&:= \lambda \trace\left(\mathbf A^T \mathbf A\right) -2\trace(\bm\Theta^T\bA)+\trace(\bm\Theta^T\bar\bA)\label{g1prooflemma}
 \end{align}
 where $\lambda:=\lambda_{\max}(\bR_A^{-1})$ and $\bm\Theta:=\lambda\bI-\bR_A^{-1}$, and
\begin{align}
f_2(\bA)&:=-\mathbf 1_M \bDelta\hadprod\bZ\log(\bA\bm\Pi^T)\mathbf 1_{NP}\label{f2_proof_lemma}\\
g_2(\bA,\bar\bA)&:=-\summ \sum_{k=1}^{NP}\delta_{mk}z_{mk}\alpha_{mkr}\log\left(\frac{a_{mr}\pi_{kr}}{\alpha_{mkr}}\right)\label{g2a}
 \end{align}
 with $\alpha_{mkr}:=\bar a_{mr'}\pi_{kr'}/ \sum_{r'=1}^R \bar a_{mr'}\pi_{kr'} .$

Hence,  properties i)-iii) will be satisfied  by the pair of  functions $g(\bA,\bar\bA)$ and $f(\bA)$  in Lemma \ref{lemma:majorizing}, as long as  they are satisfied both by the pair  in \eqref{f1prrof_lemma}-\eqref{g1prooflemma}  and that in \eqref{f2_proof_lemma}-\eqref{g2a}.

Focusing on the first pair, both functions are  separable per column of $\bA$ and $\bar\bA$, and their difference takes the form
\begin{align*}
g_1(\bA,\bar\bA)\hspace{-.05cm}-\hspace{-.05cm}f_1(\bA)\hspace{-.1cm}&=\hspace{-.1cm}\sumr[\lambda
\mathbf a_r^T\mathbf a_r -2\bm\theta_r^T\mathbf a_r+
\bm\theta_r^T\bar{\mathbf a}_r -
\bar{\mathbf a}_r^T \bR_A^{-1}\bar{\mathbf a}_r]\\
&=\sumr(\mathbf a_r -\bar{\mathbf a}_r)^T(\lambda \bI -\bR_A^{-1})(\mathbf a_r -\bar{\mathbf a}_r)
 \end{align*}
which is  positive and, together with its gradient, vanish at $\bar \bA$. This establishes that properties i)-iii) are satisfied by $g_1(\bA,\bar\bA)$ and $f_1(\bA))$,   and thus they are so for functions $g(\bA,\bar\bA)$ and $f(\bA))$ in \eqref{g_ls} and \eqref{f_ls}.

Considering  the second pair, and expanding $f_2(\bA)$ yields
\begin{align}
f_2(\bA)=-\summ \sum_{k=1}^{NP}\delta_{mk}z_{mk}\log\left(\sum_{r'=1}^R a_{mr'}\pi_{kr'}\right)\label{f2a}
\end{align}
where the logarithm can be rewritten as (see also \cite{kolda_poisson})
\begin{align}
\log\left(\sum_{r'=1}^R a_{mr'}\pi_{kr'}\right)=\log\left(\sum_{r'=1}^R\alpha_{mkr'} \frac{  a_{mr'}\pi_{kr'}}{\alpha_{mkr'}}\right)\label{log_comb_f}\\
\geq \sumr \alpha_{mkr}\log\left(\sum_{r'=1}^R\frac{ a_{mr'}\pi_{kr'}}{\alpha_{mkr}}\right)\label{log_comb_g}
\end{align}
and the inequality holds because of  the concavity of the logarithm with an argument being a convex combination with coefficients  $\{\alpha_{mkr}\}_{r=1}^R$ summing up to one.

Since substituting \eqref{log_comb_g} for \eqref{log_comb_f} in \eqref{f2a} results in \eqref{g2a}, it follows that
$g_2(\bA,\bar\bA)$ and $f_2(\bA)$ satisfy property iii). The proof is complete after evaluating at the pair of functions and their derivatives at $\bA$ to confirm that properties i) and ii)  hold too.

The minimum  $a_{g,mr}^{\star}:=t_{mr}+\sqrt{t_{mr}^2+s_{mr}}$ is obtained readily after equating to zero the derivative of the corresponding summand in \eqref{g_lsstandars}, and selecting the nonnegative root.
\end{IEEEproof}

\end{document}